\documentclass[10pt]{iopart}

\usepackage{amsfonts}
\usepackage{amssymb}
\usepackage{hyperref}
\usepackage[shortcuts]{extdash}
\usepackage{graphicx}
\usepackage{color}
\usepackage{xcolor}
\usepackage{cite}
\usepackage[mathscr]{euscript}
\usepackage{rotating}
\usepackage{siunitx}
\usepackage{tabularx}

\newcommand{\mean}[1]{\langle#1\rangle}

\hypersetup{colorlinks=true,linkcolor=magenta
  ,citecolor=magenta, urlcolor=magenta}

\begin{document} 
\flushbottom 

\title{Joint subnatural-linewidth and single-photon emission from
  resonance fluorescence} 

\author{J.~C.~{L\'{o}pez~Carre\~{n}o},$^{1,2}$
  E. Zubizarreta~Casalengua,$^1$ F.~P.~Laussy$^{2,3}$ and E.~del~Valle$^1$} 

\address{$^1$Departamento de F\'isica Te\'orica de la Materia
  Condensada, Universidad Aut\'onoma de Madrid, 28049 Madrid, Spain}

\address{$^2$Faculty of Science and
  Engineering, University of Wolverhampton, Wulfruna St, Wolverhampton
  WV1 1LY, UK}

\address{$^3$Russian Quantum Center, Novaya 100, 143025 Skolkovo,
  Moscow Region, Russia}
\ead{elena.delvalle.reboul@gmail.com}

\date{\today}

\begin{abstract}
  Resonance fluorescence---the light emitted when exciting resonantly
  a two-level system---is a popular quantum source as it seems to
  inherit its spectral properties from the driving laser and its
  statistical properties from the two-level system, thus providing a
  subnatural-linewidth single-photon source. However, these two
  qualities do not actually coexist in resonance fluorescence, since
  an optical target detecting these antibunched photons will either be
  spectrally broad itself and not benefit from the spectrally narrow
  source, or match spectrally with the source but in this case the
  antibunching will be spoiled. We first explain this failure through
  a decomposition of the field-emission and how this gets affected by
  frequency resolution. We then show how to restore the sought joint
  subnatural linewidth and antibunched properties, by interfering the
  resonance fluorescence output with a coherent beam. We finally
  discuss how the signal that is eventually generated in this way
  features a new type of quantum correlations, with a plateau of
  antibunching which suppresses much more strongly close photon
  pairs. This introduces a new concept of perfect single-photon
  source.
\end{abstract}


\section{Introduction}
\label{sec:1}

Resonance fluorescence has always been a central topic in quantum
optics, being the simplest nontrivial quantum light source: a
two-level system driven coherently close to, or at, its
resonance~\cite{heitler_book44a,mollow69a,wu75a,agarwal76a,
  kimble76a, cohentannoudji77a, grove77a, kimble77a, dagenais78a,
  knight78a}. Early on, it has been recognised as a single-photon
source (SPS) that should exhibit perfect antibunching, that is, a
complete suppression of photon coincidences. Intuitively, this is
because no photon can be emitted (or detected by an ideal detector) at
the same time as another one, due to the finite reloading time of the
system after an emission. This experimental observation made resonance
fluorescence, in fact, the first system to fully prove the
quantization of light~\cite{kimble77a}, by violating the classical
Cauchy-Schwartz inequality for the intensity-intensity correlations in
time.  This has since been tested and confirmed throughout the history
of the field in a variety of platforms~\cite{apanasevich79a,
  cohentannoudji79a, mandel79a, singh83a, carmichael85a,
  kozlovskii99a, astafiev10a, verma11a}. It also created an obvious
incentive of perfecting this source of single-photons for
applications, since a SPS is a crucial component of quantum technology
in most platforms, including cold atoms~\cite{itano88a,
  grangier86b,rempe90a},
ions~\cite{diedrich87a,bergquist86a,schubert92a},
molecules~\cite{kask85a,basche92a,treussart01a,lounis00a},
semiconductor quantum dots~\cite{michler00a,lounis00b, santori01a,
  zwiller02b, sebald02a,santori02a, pelton02a,yuan02a, gerardot05a},
superconducting circuits~\cite{astafiev10a,bozyigit11a, lang13a,
  hoi13a, gu17a}, nitrogen
vacancies~\cite{kurtsiefer00a,brouri00a,messin01b}, and still others.
Recent years have been particularly fruitful towards the
implementation of an ideal SPS ripe for commercial development and
industrial applications~\cite{kuhlmann15a,somaschi16a,ding16a,wang16a,
  kim16b,daveau17a,grange17a}.  In this respect, resonance
fluorescence appears to be among the best contenders.  Together with
its sub-Poissonian statistics, it also has a very strong emission rate
thanks to the efficient coherent driving, and, in contrast to
incoherent driving that results in power broadening, it can be
operated in the so-called \emph{Heitler regime}~\cite{heitler_book44a}
where its spectral width is actually narrower than the natural
linewidth of the emitter, being instead given by the driving
laser. This led to the claim of the emission as an elastic scattering
(i.e., \emph{Rayleigh}) peak, which retains the coherence as well as
spectral width of the laser~\cite{gibbs76a, hartig76a}, and the
antibunching of the two-level system~\cite{hoffges97a}.

Resonance fluorescence is therefore a precious resource, since all
these three attributes are precisely those demanded by the prospective
quantum circuits for the technology of tomorrow: antibunching to deal
with quantum states, brightness to provide a strong signal and narrow
spectral-width to have indistinguishable photons.  These qualities
were first explored with a single trapped ion~\cite{hoffges97a} and
more recently exploited with a single semiconductor quantum
dot~\cite{nguyen11a,matthiesen12a,unsleber16a,loredo16a,somaschi16a,he17a}
which is still under active development. All these studies follow a
similar trend: they analyse spectral properties with the best
available spectral resolution on the one hand, and then the
statistical properties (the second-order correlation function) with
the best available temporal resolution on the other hand. These
constitute two different experiments, providing excellent results in
both cases and seemingly fulfilling the ideal scenario we have just
described: perfect antibunching of spectrally narrow sources. However,
one should contrast these qualities \emph{together}, that is to say,
simultaneously. One is ultimately interested not in how well the
source performs when considering one aspect or the other in isolation,
but how an optical target that is excited by the source will
``perceive'' these photons. Such a target will have a spectral
width~$\Gamma$ and couple to the source accordingly, preventing it to
see the photon statistics with an independent time resolution, that is
needed to extract the best antibunching. Therefore, to properly
describe the SPS, one needs to study the spectral and statistical
properties of resonance fluorescence as detected in one and the same
experimental setup, including the Heisenberg time and frequency
uncertainties.

\begin{figure}[t]
  \centering 
  \includegraphics[width=0.6\linewidth]{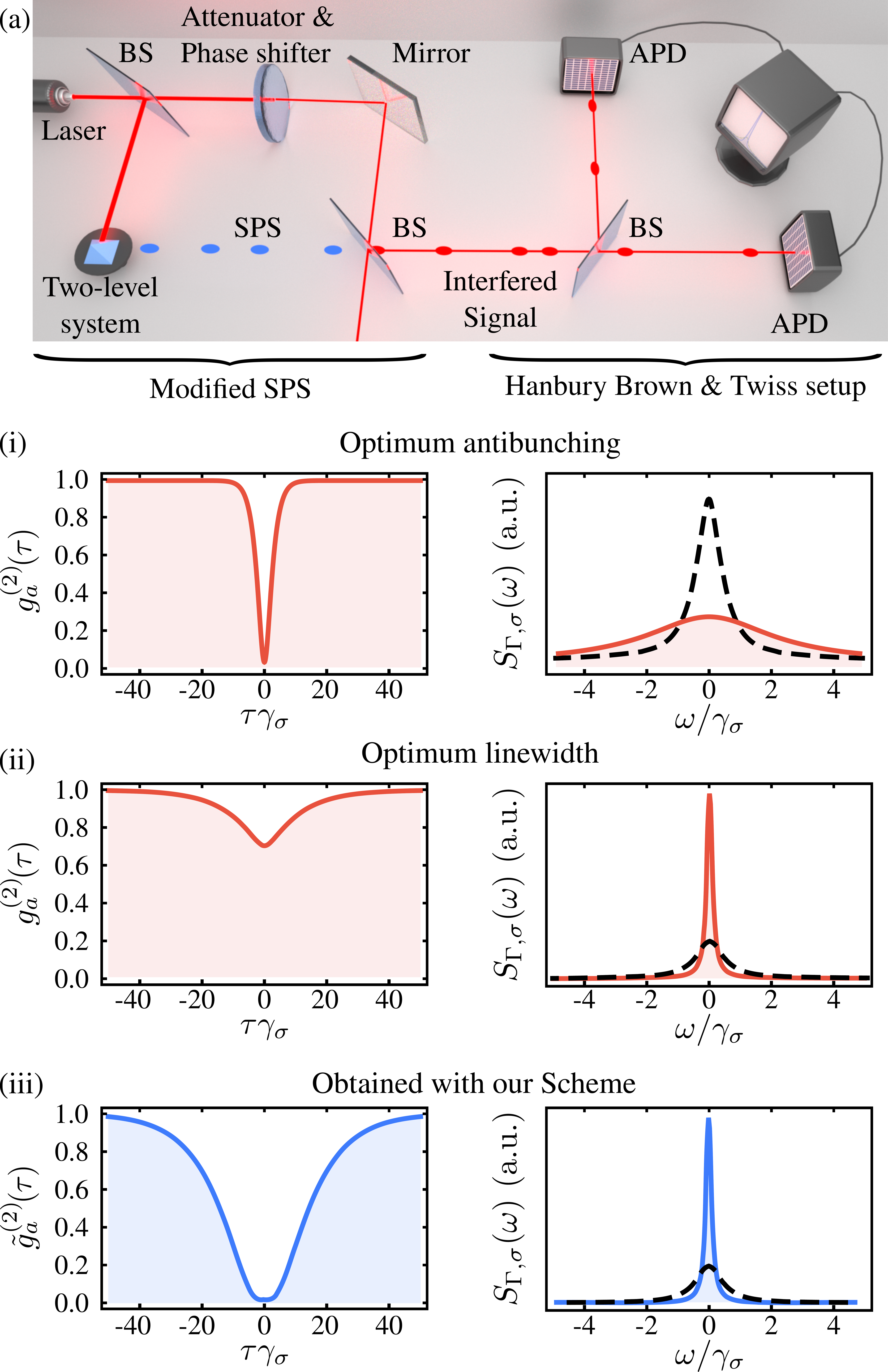}
  \caption{(a)~Scheme of our proposed setup to generate a
    single-photon source for which one can simultaneously measure, in
    the same experiment and with both time- and frequency-resolving
    detectors, a narrow spectrum of emission and perfect
    antibunching. From left to right: Part of the excitation laser
    (red beam) is attenuated and $\pi$-phase shifted, to later
    interfere with the resonance fluorescence signal (blue dots). The
    right-hand side of the table represents a standard Hanbury-Brown
    Twiss setup to measure the second-order correlation of the total
    signal. (i) Using spectrally wide detectors to measure
    antibunching broadens the spectrum of emission (solid red) as
    compared to the natural linewidth of the emitter (dashed
    black). (ii) Using spectrally narrow detectors resolves well in
    frequency but spoils the antibunching. (iii) Using the scheme
    in~(a) with narrow detectors, we can have simultaneously perfect
    antibunching (iii) and a narrow spectrum (solid blue).}
  \label{fig:1}
\end{figure}

Doing so, we find that for resonance fluorescence, subnatural
linewidth of the emission is not compatible with a simultaneous strong
antibunching. The observed (or detected) linewidth of the Rayleigh
peak is broadened by the spectral resolution~$\Gamma$. Keeping this
broadening below the natural two-level system decay
rate~$\gamma_\sigma$ spoils the antibunching and brings the statistics
to the Poissonian limit. The expression for the filtered (or
  convoluted with the detector) second-order correlation function of
resonance fluorescence at low driving is indeed known to
be~$g_a^{(2)}=[\gamma_\sigma/(\gamma_\sigma+\Gamma)]^2$, which goes
to~1 as $\Gamma\rightarrow 0$~\cite{gonzaleztudela13a}. Antibunching
is thus washed out by the large detector time
uncertainty~$1/\Gamma$. This incompatibility is shown for our problem
at hand in rows~(i-ii) of Fig.~\ref{fig:1}.

Although it does not work with resonance fluorescence per se, the
intuition of the pioneering experiments~\cite{nguyen11a,matthiesen12a}
to realise a subnatural-width antibunched source (implying,
\emph{simultaneously}), is not forbidden on fundamental grounds: one
can imagine a source as spectrally narrow and antibunched as one
wishes, merely by changing the
timescale~($1/\gamma_\sigma$). There is therefore no a-priori
reason why the initial claim could not be realised one way or the
other.

In this text, we present a scheme to do that, that is, to provide
perfect antibunching from resonance fluorescence without renouncing to
subnatural linewidth. This is achieved, in contrast to previous works,
so that the \emph{same} detection setup measures \emph{simultaneously}
these two quantities: antibunching and narrow spectral width. We base
our protocol on the understanding of such perfect antibunching as the
result of destructive interference between the coherent and incoherent
fractions of the emission: the coherently scattered photons and those
that are absorbed and re-emitted~\cite{carmichael85a}. The detector
can then be seen as a filter that breaks the equilibrium between these
two fractions, absorbing more coherent than incoherent light. We can
thus restore this equilibrium since coherent light is easy to
control. We propose to do so with a setup such as the one sketched in
Fig.~\ref{fig:1}(a), where the coherent fraction in the resonance
fluorescence signal is reduced by making it interfere with an external
$\pi$~phase-shifted laser beam, attenuated to the right proportion for
the compensation to be perfect. We provide the exact (analytical)
condition for this to occur as well as a full analysis of the
spectral, statistical and intensity properties in terms of all the
relevant parameters of the problem. We also show that, in fact, such a
source goes even further and behaves more closely to an ideal SPS than
would resonance fluorescence alone operating in a different timescale.

The rest of the paper is organised as follows: In Sec.~\ref{sec:2}, we
review the spectral and statistical properties of resonance
fluorescence for ideal and realistic detectors, introducing the
theoretical formalism as we do so, and we show how antibunching can be
interpreted in terms of coherent and incoherent contributions to the
second-order correlation function. In Sec.~\ref{sec:3}, we present the
setup to obtain perfect antibunching and high frequency resolution
when considering realistic and simultaneous measurement of statistical
and spectral properties, based on a complete theoretical
description. We provide analytical expressions for the condition to be
fulfilled, that could guide its experimental realisation. In
Sec.~\ref{sec:4}, we further analyse other important quantities to
characterise the system, namely, the coherence time of the
second-order correlation function and the emission rate of the
source. Finally, in Sec.~\ref{sec:5}, we conclude.

\section{Antibunching in resonance fluorescence and the impact of detection}
\label{sec:2}

We consider the low driving regime of resonance fluorescence, or
so-called \emph{Heitler regime}~\cite{heitler_book44a}. In this
scenario, the emitter is modelled as a two-level system with
annihilation operator~$\sigma$ and is driven coherently with a weak
laser of intensity~$\Omega_\sigma$. We consider the laser exactly at
resonance with the two-level transition for simplicity but everything
can be easily generalised to the close-to-resonance case by adding a
detuning parameter. Importantly, we take into account the physical
detection of resonance fluorescence. This is a central point of our
approach as it allows us to consider the physical, self-consistent and
complete description of the source. In particular, this accounts for
the uncertainty in time and frequency of the detected
photons~\cite{eberly77a}. Technically, this involves the integration
of the convolution between the observable and a filtering function,
which becomes exponentially difficult as the number of photons
involved in the observable increases~\cite{knoll86a,knoll90a}. Such a
difficulty can be overcome if the detectors are considered as physical
passive objects that receive the emission of the quantum source
without disturbing it. This can be obtained when the detectors are
described as harmonic oscillators that couple to the source either in
the limit of vanishing coupling~\cite{delvalle12a} or through the
so-called~\emph{cascaded coupling}~\cite{gardiner_book00a}. In either
of these equivalent methods~\cite{lopezcarreno18a}, the
excitation is allowed to go from the quantum source to the detector
while the feedback in the opposite direction is suppressed. Following
these ideas, our detector is therefore considered as an harmonic
oscillator, with bosonic annihilation operator~$a$, and the full and
self-consistent description of resonance fluorescence becomes an easy
theoretical problem again. Indeed, the master equation describing this
complete system is given by~(we take~$\hbar=1$ from now on):
\begin{equation}
      \label{eq:TueNov14183943CET2017}
      \partial_t \rho = i[\rho,H] +
      \frac{\gamma_\sigma}{2}\mathcal{L}_\sigma \rho +
      \frac{\Gamma}{2}\mathcal{L}_a\rho\,.
\end{equation}
The dissipation
term~$\mathcal{L}_c=2c\rho c^\dagger-c^\dagger c\rho-\rho c^\dagger c$
is in the Lindblad form, with $\gamma_\sigma$ and $\Gamma$ being the
decay rates of the two-level system and the detector,
respectively. The parameter $\Gamma$ provides the spectral width of
the detector and its inverse, $1/\Gamma$, thus gives the temporal
uncertainty of the detector. The Hamiltonian,
$H=\Omega_\sigma (\sigma+\sigma^\dagger)+ g(a^\dagger\sigma +
\sigma^\dagger a)$, describes the laser driving the two-level system
(with a parameter~$\Omega_\sigma$ that we consider to be real without
loss of generality) and its coupling to the detector is taken as~$g$
(also real). We set the detector at resonance with both the laser and
the two-level system.

One of the central quantities in this work is the second-order
correlation function~\cite{glauber63a}, typically defined, for a
source with operator~$s$ in the steady state, as:
\begin{equation}
  \label{eq:MonNov27154304CET2017}
  g_s^{(2)}(\tau)=\lim_{t\rightarrow \infty}\frac{\mean{s^\dagger (t)
      (s^\dagger s )(t+\tau) s (t)}}{[\mean{s^\dagger s}(t)]^2} =
  \frac{\mean{s^\dagger  (s^\dagger s)(\tau) s}}{\mean{s^\dagger s}^2} \,.
\end{equation}
We omit the time $t$ in all expressions, which we consider to be large
enough for the system to have reached the steady state. When the
delay~$\tau$ is omitted as well, it is implicitly assumed to be zero:
$g_s^{(2)}=g_s^{(2)}(\tau=0)$, which describes coincidences. We will also
be considering the $N$th-order correlation functions, but then always
at zero time-delay:
$g_s^{(N)}= \mean{s^{\dagger N} s^N}/\mean{s^\dagger s}^N$.

Let us start by reviewing the spectral properties of this system with
perfect frequency resolution~\cite{mollow69a,loudon_book00a}. The
details of the derivation can be found in~\ref{app:2}. The
normalised steady-state spectrum of emission in the low driving
regime, $\Omega_\sigma\ll \gamma_\sigma$, formally defined in
Eq.~(\ref{eq:TueNov28145129CET2017}), reads
\begin{equation} 
\label{eq:MonJul21132824CEST2008}
S_\sigma(\omega)=(1-K_2)\delta(\omega)+K_2\frac{1}{\pi}\frac{\frac{\gamma_\sigma}{2}}{\big(\frac{\gamma_\sigma}{2}\big)^2+\omega^2}\,,
\end{equation} 
where $K_2$ is given by, up to second order in the driving,
$K_2=8\Omega_\sigma^2/\gamma_\sigma^2$.  This is simply the
superposition of a delta and a Lorentizan peaks, both centered at the
laser frequency (at zero), with no width and $\gamma_\sigma$-width,
respectively. The delta function term is the Rayleigh peak attributed
to the elastic scattering of the laser photons by the two-level system
while the Lorentzian term comes from the actual two-photon excitation
and re-emission~\cite{dalibard83a}. Note that in the linear regime and
particularly in the limit $\Omega_\sigma\rightarrow 0$ and excluding
second order terms (which involve two-photon states in the detector),
the spectrum of emission reduces to the delta function. That is, if
one is interested in the spectral density of isolated one-photon
events only, regardless of their time of arrival or their relation to
other photons, the source is effectively providing photons as
spectrally narrow as the laser (here infinitely narrow making the
source perfectly monochromatic). However, if such photons are to be
used in temporal relation with others, such as when considering their
antibunching properties, then the second order part of the spectrum
must be taken into account. By having the frequency resolution
below the natural emitter linewidth (in order to maintain a narrow
spectrum to first order) one filters out part of the incoherent
spectrum which determines its statistics. On the other hand,
increasing the frequency resolution in order to increase temporal
precision, broadens the spectrum. As a result, resolving antibunching
spoils the subnatural linewidth of the source, and vice-versa. To make
this important point more quantitative, let us consider $g_a^{(N)}$
the $N$th-order correlation function of resonance fluorescence as
measured by a detector with both frequency and time resolution (set at
resonance with the source). The expressions for a general laser
driving strength exist but are bulky~(see, for instance, the
case~$N=2$ in Eq.~(19b) of Ref.~\cite{lopezcarreno16a}). Here,
since we are interested in the Heitler regime, it is enough to expand
these expressions to the lowest order in the driving, which is, for
$\mean{a^{\dagger N}a^N}$, to order $O(\Omega_\sigma^{2N})$, as shown
in~\ref{app:1} with $\Omega_a=0$. This allows us to
generalise to all orders the expression for the correlations, that
simply reduces to (for $N\geq 2$):
\begin{equation}
  \label{eq:WedNov15091914CET2017}
  g_a^{(N)} = \prod_{k=1}^{N-1}\frac{\gamma_\sigma^2}{(\gamma_\sigma+k
    \Gamma)^2}\,.
\end{equation}
We have already discussed the case~$N=2$ above.  As expected, when
$\Gamma\rightarrow \infty$, this expression recovers the perfect
antibunching of the source itself, i.e., when the full emission is
being detected without any frequency
resolution:~$\lim_{\Gamma\rightarrow \infty}g^{(N)}_a =g^{(N)}_\sigma
=0$~\cite{delvalle12a}. In the opposite limit of narrow frequency
filtering, the result for a coherent field is
obtained:~$\lim_{\Gamma\rightarrow 0}g^{(N)}_a =1$ for all~$N$. With
the present semi-classical model for the laser, which has zero
linewidth (perfect first-order coherence), we do not recover the
expected thermal value for photons of completely undetermined time of
emission, i.e., $\lim_{\Gamma\rightarrow 0}g^{(N)}_a \neq N!$, because
it is impossible to filter inside the laser
width~\cite{gonzaleztudela13a}. For a general intermediate $\Gamma$,
the perfect antibunching needed for quantum applications, is spoiled:
$0<g_a^{(N)}\leq 1$. For instance, when filtering at the natural
linewidth of the emitter~$\Gamma=\gamma_\sigma$, we obtain a reduction
of 25\% in the antibunching~($g_a^{(2)} = 1/4$) and
$\Gamma=\gamma_\sigma/3$ leads to~$g_a^{(2)} =0.56$. As a consequence,
making use of the subnatural spectral width of such a
SPS~\cite{matthiesen12a}, which implies detecting its photon with some
accuracy in time and frequency, or coupling its output light to an
optical element with $\Gamma<\gamma_\sigma$, spoils its statistical
properties. In summary, subnatural linewidth and antibunching are in
contradiction for resonance fluorescence in its bare form. The system
emits photons which one can choose to see, depending on the detection
scheme, with the properties of the driving laser or of the emitter,
but not simultaneously.

\begin{figure}[t]
  \centering
  \includegraphics[width=0.6\linewidth]{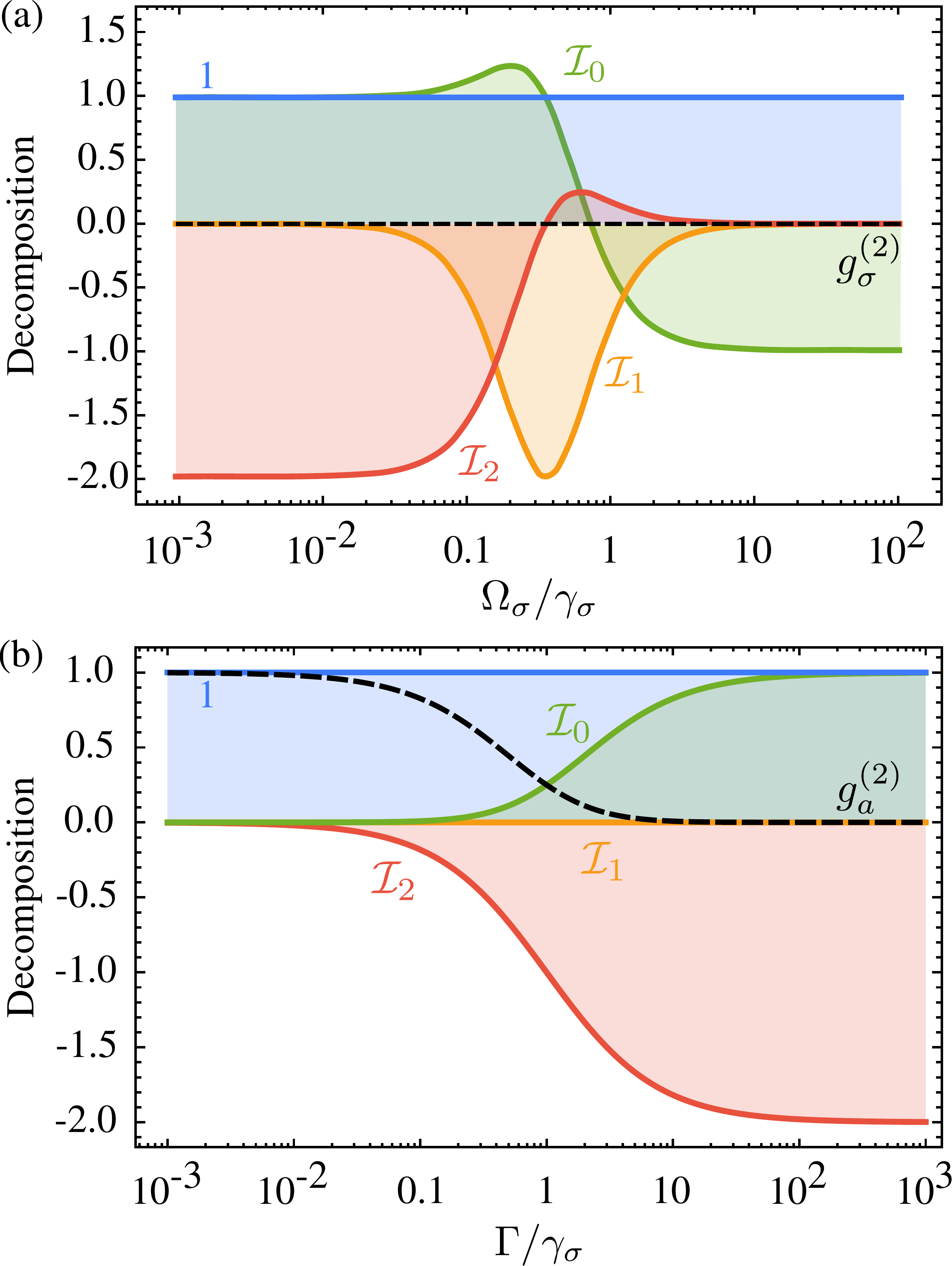}
  \caption{(a)~Second-order correlation function of the emission from
    a two-level system~(dashed black line) and its decomposition
    Eq.~(\ref{eq:WedNov15142959CET2017}) into the four components
    (solid colored lines) given by
    Eqs.~(\ref{eq:Mon12Feb162245GMT2018}-\ref{eq:Mon12Feb162717GMT2018}),
    as a function of the laser excitation. Regardless of the driving
    regime, the total emission fulfills~$g_\sigma^{(2)}=0$. In the
    Heitler regime, on the left hand side, this is due to a
    destructive interference at the two-photon level between the
    coherent and incoherent (squeezed) components of resonance
    fluorescence. In the strong driving regime, right-hand side,
    perfect antibunching is due to the dominating sub-Poissonian
    fluctuations. (b)~The same decomposition but now for the filtered
    emission~$g_a^{(2)}$ and as a function of~$\Gamma$ in the Heitler
    regime ($\Omega_\sigma=10^{-3} \gamma_\sigma$). In this case,
    antibunching gets spoiled as the frequency-resolution is increased
    by filtering, leading to an imperfect compensation of the
    components. This can however be restored with an external laser.}
  \label{fig:WedNov15170803CET2017}
\end{figure}

In order to address this discouraging issue, let us dive deeper into
the mechanism that yields antibunching in resonance fluorescence when
integrating over its full spectrum, with $\Gamma\rightarrow
\infty$. This can be understood in terms of interfering fields: it has
indeed been long known that the emission of a coherently driven
two-level system can be expressed as a superposition of a coherent and
a squeezed incoherent field~\cite{mandel82a,carmichael85a}. We apply a
mean field procedure and write~$\sigma=\alpha+d$, with a mean coherent
field $\alpha=\mean{\sigma}$ and $d$ an operator for the quantum
fluctuations characterised by $\mean{d}=0$. The coherent field is the
one that gives rise to the delta function in the spectrum of
emission~(\ref{eq:MonJul21132824CEST2008}) while the Fourier transform
of $\mean{d^\dagger d(\tau)}$ provides the Lorentzian peak, i.e., the
incoherent part of the spectrum, that transforms into the Mollow
triplet when the driving increases~\cite{mollow69a}. Working out
$g_\sigma^{(2)}$ in term of this decomposition results in four
contributions:
\begin{equation}
  \label{eq:WedNov15142959CET2017}
  g_\sigma^{(2)} = 1+\mathcal{I}_0+\mathcal{I}_1+\mathcal{I}_2 \,,
\end{equation}
that grow as powers of $\alpha$:
\numparts
\begin{eqnarray}
    \mathcal{I}_0 &=& \frac{\mean{d^{\dagger 2}d^2}-\mean{d^\dagger
          d}^2}{\mean{n_\sigma}^2}= |\alpha|^2 \frac{6 \mean{n_\sigma} - 4  |\alpha|^2}{\mean{n_\sigma}^2}-1\,,\label{eq:Mon12Feb162245GMT2018}\\
    \mathcal{I}_1 &=& 4\frac{\Re (\alpha^* \mean{d^\dagger d^2})}{\mean{n_\sigma}^2}= 8 |\alpha|^2 \frac{|\alpha|^2- \mean{n_\sigma}}{\mean{n_\sigma}^2  }\,,\label{eq:Mon12Feb162452GMT2018}\\
    \mathcal{I}_2 &=& 2\frac{|\alpha|^2 \mean{d^\dagger d} +\Re(
          \alpha^{*2}\mean{d^2} )}{\mean{n_\sigma}^2}=2
                    |\alpha|^2\frac{\mean{n_\sigma} -  2
                    |\alpha|^2}{\mean{n_\sigma}^2}\,.
                    \label{eq:Mon12Feb162717GMT2018}
\end{eqnarray}
  \endnumparts
From the derivation in~\ref{app:2}, we can further
substitute~$\alpha = -2i
\Omega_\sigma\gamma_\sigma/(\gamma_\sigma^2+8\Omega_\sigma^2)$
and~$\mean{n_\sigma}=\mean{\sigma^\dagger
  \sigma}=4\Omega_\sigma^2/(\gamma_\sigma^2+8\Omega_\sigma^2)$. This
decomposition is what one would obtain when performing a $g^{(2)}$
measurement on the output of a beam splitter, that would have $\sigma$
as the associated output arm operator, with input fields~$\alpha$ and
$d$. This is the well known homodyne measurement, first suggested by
Vogel~\cite{vogel91a,vogel95a} to analyse the squeezing properties of
signal~$d$ thanks to the controlled variation of a local
oscillator~$\alpha$. The numerator of~$\mathcal{I}_0$ in the left-hand
side of Eq.~(\ref{eq:Mon12Feb162245GMT2018}) is the normally ordered
variance of the fluctuation intensity, i.e.,
$\mean{{:}(\Delta n_d)^2{:}}=\mean{{:}n_d^2{:}}-\mean{n_d}^2$ with
$n_d=d^\dagger d$ and $\Delta n_d=n_d-\mean{n_d}$. Therefore, having
$\mathcal{I}_0<0$ indicates sub-Poissonian statistics of the fluctuations,
which, in turn, contributes to the sub-Poissonian statistics of the
total field~$\sigma$. The numerator of~$\mathcal{I}_1$ in
Eq.~(\ref{eq:Mon12Feb162452GMT2018}) represents the normally ordered
correlation between the fluctuation field-strength and intensity,
$\mean{d^\dagger d^2}=\mean{{:}\Delta d \, \Delta n_d{:}}$, which have
been referred to as \emph{anomalous
  moments}~\cite{vogel91a,vogel95a}. A squeezed-coherent state has
such correlations. The numerator of the last component, $\mathcal{I}_2$, in
Eq.~(\ref{eq:Mon12Feb162717GMT2018}), can be written in terms of one
of the fluctuation quadrature~$X=(d + d^\dagger)/2$, in the following
way: $ 4 |\alpha|^2 \left( \mean{{:}X^2{:}} - |\alpha|^2 \right)$. If
this is negative, there is some quadrature squeezing.
	
The four terms of this decomposition for $g_\sigma^{(2)}$ are shown in
Fig.~\ref{fig:WedNov15170803CET2017}(a), as a function of the
intensity of the driving laser. They always compensate exactly and the
final result is, of course, the perfect sub-Poissonian statistics of
the two-level system emission. However, as is clear in the figure,
this compensation occurs in different ways depending on the driving
regime~\cite{carmichael85a}:

\begin{itemize}
\item In the region of large driving
  ($\Omega_\sigma \gg \gamma_\sigma$), where the spectrum of the
  emitter displays a Mollow triplet, we have that $\mathcal{I}_0=-1$
  with $\mathcal{I}_1=\mathcal{I}_2=0$ meaning that antibunching
  appears solely due to the sub-Poissonian statistics of the
  fluctuations, that dominate over the vanishing coherent component
  $\lim_{\Omega_\sigma\rightarrow \infty}\alpha=0$ (and therefore
  $d\rightarrow \sigma$).
\item In the intermediate driving region
  ($\Omega_\sigma \sim \gamma_\sigma$), it is $\mathcal{I}_1<0$ that almost
  fully compensates the positive contributions of $1+\mathcal{I}_0$. This is
  where $\mathcal{I}_0$ changes sign and the fluctuations become
  super-Poissonian.
\item In the Heitler regime~($\Omega_\sigma \ll \gamma_\sigma$) that
  interests us more particularly, it is fluctuation squeezing that
  plays a major role, being this time the one responsible for
  antibunching, $\mathcal{I}_2=-(1+\mathcal{I}_0)=-2$.
\end{itemize}
Note that in the Heitler regime, $\mathcal{I}_1$ vanishes
again. Consequently, resonance fluorescence reaches its maximum
squeezing also in this region, an effect that has been confirmed in
the emission from ensembles of
atoms~\cite{slusher85a,mccormick07a,mccormick08a,raizen87a,lu98a} and
recently also from single atoms~\cite{ourjoumtsev11a} and quantum
dots~\cite{schulte15a}. Another way to understand the origin of
antibunching in this region is as an interference between the coherent
and incoherent parts of the
emission~(c.f.~Eq.~(\ref{eq:MonJul21132824CEST2008})), since those
terms which are either purely coherent (1) or purely incoherent
($\mathcal{I}_0$), are fully compensated by the 50\%-50\% mixed
one,~$\mathcal{I}_2$.

One can also compute the decomposition for the filtered second-order
correlation function~$g^{(2)}_a$ by applying again
Eqs.~(\ref{eq:Mon12Feb162245GMT2018}-\ref{eq:Mon12Feb162717GMT2018}),
now with the detector field operators, that is with
$a\rightarrow \sigma$, $\mean{n_\sigma}\rightarrow \mean{n_a}$ and
$\alpha=\mean{a}$. This is shown in
Fig.~\ref{fig:WedNov15170803CET2017}(b) for the Heitler regime as a
function of the filter width~$\Gamma$. One can see how, with
filtering, or equivalently when detection is taken into account, the
terms no longer exactly compensate each other and their sum do not add
up to exactly $g_a^{(2)}=0$. Speaking in spectral terms, this is
because the filter is leaving out some of the incoherent part that
should compensate for the fixed coherent one (the delta function is
always fully included in the convolution with the filter centered
at~$\omega_\mathrm{L}=0$). In the Heitler regime, this is clear when
$\Gamma<\gamma_\sigma$, since $\gamma_\sigma$ is the width of the
(incoherent) Lorentzian peak, as shown in
Eq.~(\ref{eq:MonJul21132824CEST2008}).

\section{Destructive $N$-photon interference and antibunching
  restoration}
\label{sec:3}

The decomposition of the filtered second-order correlation
function~$g_a^{(2)}$ outlined above allows us to determine what is
missing in terms of coherent and/or incoherent fractions to produce
the perfect antibunching. Since the compensation comes, in part, from
a coherent field, and such a field is easy to produce and control in
the laboratory, one can actually \emph{restore} full antibunching by
superimposing to the filtered resonance fluorescence an external
coherent field~$\beta$, making them interfere at a beam splitter, and
collecting the new signal for further use or analysis. We can find
theoretically the value of $\beta$ that ensures that the resulting
total field, $s=t\,\sigma +r\,\beta$ (with $t$ and~$r$ the
transmission and reflection coefficients of the beam splitter,
respectively, taken real and such that $r^2+t^2=1$), although it has
been filtered, still produces perfect antibunching at the output. We
will call~$\tilde g_a^{(2)}$ the second-order correlation function of
this filtered signal that is interfered with a correcting external
coherent beam, and proceed to show how to cancel it despite the
filtering. This, in effect, realises the previously claimed subnatural
linewidth single-photon source~\cite{nguyen11a,matthiesen12a}. This
becomes possible because the source is not a passive object anymore,
that relates time and frequency of its emission merely through the
Fourier transform, but includes a dynamical element. We will see in
the following that, as a consequence, our source even achieves more
than joint subnatural linewidth and antibunching.

The principle for antibunching restoration is simple. Since the
filtering reduces the incoherent fraction, $\beta$ should lower
(proportionally) the coherent fraction. This is possible for two
coherent fields by destructive interferences. That is, given that at
resonance $\alpha=-i|\alpha|$, we should find a $\beta$ of the
form~$\beta=i|\beta|$ such that the total mean field is reduced to
$-i(t|\alpha|-r|\beta|)$. Out of resonance, both $\alpha$ and $\beta$
have imaginary and real parts but the same idea would apply. This
protocol and the condition for the external $\beta$-field are one of
the chief results of this text. We now proceed to describe a possible
setup to realise this interference and a theoretical model that
provides an exact analytical condition.

The simplest and most reliable way to interfere resonance fluorescence
with a controlled coherent field is to divert some light from the
laser that excites the two-level system in the first place. In this
way, one works with the same coherence time of the driving laser and
should be immune to slow fluctuations.  A possible setup is sketched
in Fig.~\ref{fig:1}(a): The laser beam passes through a first beam
splitter~\footnote{We have not taken into account the first beam
  splitter of Fig.~\ref{fig:1} in the calculations for simplicity,
  but, assuming it is balanced (50:50), doing so would simply rescale
  the original driving to $2\Omega_\sigma$ in order to obtain the
  results in the manuscript.}, that redirects part of it to the
two-level system on one output arm and to an attenuator and a phase
shifter on the other output arm. The emission of the two-level system
($\sigma$) and the attenuated laser ($\beta$) are admixed at a second
beam splitter. The output constitutes our new antibunched source
$s=t\,\sigma +r\,\beta$, which can be further analysed,
measuring, for instance, its second-order correlation function in a
Hanbury--Brown Twiss setup, as depicted in the figure.

In the theoretical description, we include the detectors in the
dynamics to now receive simultaneously the attenuated laser and the
emission of the emitter~$s=t\,\sigma+r\,\beta$. This is modeled by
adding a coherent driving term to the detector~$H_a =
i\Omega_a(a^\dagger -a)$, substituting $H\rightarrow H+H_a$ in
Eq.~(\ref{eq:TueNov14183943CET2017}), with $\Omega_a\in
\mathbb{R}$. Note that the phase of the detector driving is fixed to
$\beta=i \Omega_a/g$ and the resulting Hamiltonian is
then~$H+H_a=\Omega_\sigma \sigma +g(t \sigma + ir \Omega_a/g)a^\dagger
+ \mathrm{h.\,c.}$ In this way, the detector is effectively performing
the described homodyne procedure between the light emitted by the
two-level system and a coherent field with amplitude $r\Omega_a/(tg)$.
Furthermore, our model describes detection self-consistently and
allows us to study the joint dynamical properties (in both time and
frequency) of the light produced by the superposition. Since we are
interested in the low driving limit, we express $\Omega_a$ in terms of
$\Omega_\sigma$ through a new dimensionless parameter:
\begin{equation}
  \label{eq:Mon12Feb215757GMT2018}
  \mathcal{F}=\frac{r}{t}\frac{\gamma_\sigma \Omega_a}{g \Omega_\sigma}\,.
\end{equation}
We also define~$\mathcal{F}'=\mathcal{F}t/r$, which absorbs the
dependence on the transmission and reflection parameters of the beam
splitter. We take both $\mathcal{F}$ and $\mathcal{F}'$ to be real and
positive. With these definitions, $\beta=i
\Omega_\sigma\mathcal{F'}/\gamma_\sigma$ and it is clear that
$100\mathcal{F'}$ is then the percentage of the laser intensity that
finally interferes with resonance fluorescence while $\mathcal{F}$ is
the fraction that is needed to attenuate the laser for the
compensation to be effective. The total mean field of the signal that
exits the beam splitter towards detection (the right-hand side in
Fig.~\ref{fig:1}(a)) reads $\mean{s}=t \alpha+r\beta=-i\Omega_\sigma
t(2-\mathcal{F})/\gamma_\sigma$.

Next, we solve the new master equation in the Heitler regime,
following the procedure in~\ref{app:1}, we find that the
detected~$N$th-order correlation function is
\begin{equation}
    \label{eq:SunNov19161840CET2017}
    \tilde g_a^{(N)} =g_a^{(N)} \left[ \frac{\sum_{k=0}^{N}
        {N \choose k}2^k
        (\mathcal{-F'})^{N-k}\prod_{\lambda=1}^{N-k}[1+(N-\lambda)\Gamma/\gamma_\sigma]}{(2-\mathcal{F'})^{N}}\right]^{2} \,,
\end{equation}
where $\tilde g_a^{(N)}$ is for the compensated signal and $g_a^{(N)}$
is given by Eq.~(\ref{eq:WedNov15091914CET2017}). Note that
all $\tilde g_a^{(N)}$ have a divergence at $\mathcal{F'}=2$,
independently of the filtering parameter $\Gamma/\gamma_\sigma$. This
is another type of interference related to superbunching that lies
beyond the scope of the present analysis and that is discussed
elsewhere~\cite{arxiv_zubizarretacasalengua18a}. For~$N=2$ the
correlation in Eq.~(\ref{eq:SunNov19161840CET2017}) simplifies to
\begin{equation}
  \label{eq:SunNov19163153CET2017}
  \tilde g^{(2)}_a=\left[
    \frac{4\gamma_\sigma-(4-\mathcal{F'})\mathcal{F'}
      (\gamma_\sigma+\Gamma)}{(2-\mathcal{F'})^2(\gamma_\sigma+\Gamma)}
    \right]^2\,,
\end{equation}
which becomes exactly zero when the attenuation factor takes the two
values
\begin{equation}
  \label{eq:WedNov15121440CET2017}
  \mathcal{F}_{2,\pm}'=2 \left(1\pm
  \sqrt{\frac{\Gamma}{\Gamma+\gamma_\sigma}}\right)\,.
\end{equation}
This result is valid to first (leading) order
in~$\Omega_\sigma$, meaning that when the
condition~(\ref{eq:WedNov15121440CET2017}) is fullfilled,
$\tilde g^{(2)}_a=0$ with deviations due to higher-order terms in the
driving only, so remaining extremely small. The antibunching becomes
``exactly zero'' only in the limit of vanishing driving.  In fact,
would~$\tilde g^{(2)}_a$ be exactly zero, then also all higher-order
terms would satisfy $\tilde g^{(N)}_a=0$
for~$N\ge 2$~\cite{zubizarretacasalengua17a} and provide the ultimate,
perfect single-photon source that emits a Fock state of a single
photon, so with vanishing signal over time. In the Heitler regime but
with a finite signal, the antibunching remains so small as to be well
approximated by zero on the figures, in contrast to normal resonance
fluorescence.

Note that the protocol we have just outlined becomes meaningless for
two extreme cases: in the limit of broad filters, where we recover
perfect antibunching without the interference,
$\lim_{\Gamma\rightarrow \infty} \mathcal{F}_{2,-}' = 0$, and in the
limit of vanishingly narrow filters, where
$\lim_{\Gamma\rightarrow 0} \mathcal{F}_{2,\pm}' = 2$ and
$\tilde g_a^{(N)}$ diverges.  The latter case means that if the filter
is very narrow, compensating for the loss of the incoherent component
becomes impossible, as one ends up removing completely the coherent
component with no signal left. One can nevertheless reduce the
linewidth by over an order of magnitude as compared to the natural
linewidth of the emitter, which amply qualifies as a subnatural
linewidth.

The two solutions in Eq.~(\ref{eq:WedNov15121440CET2017}) correspond
to two different mean fields that, despite having different phases,
lead to the same intensity in the interference signal,
$\mean{s_\pm}=t\alpha+r\beta_\pm=\pm 2i\Omega_\sigma
t/\gamma_\sigma\sqrt{\Gamma/(\Gamma+\gamma_\sigma)}$, and, therefore,
both successfully compensate for the incoherent component, recovering
perfect antibunching for any given realistic detection
resolution~$\Gamma$. Nevertheless, they are of a very different
character: $\mathcal{F}_{2,+}'$ changes the phase of the original mean
field, from $\alpha =-i|\alpha|$ to $\mean{s_+}=i|t\alpha+r\beta_+|$,
while $\mathcal{F}_{2,-}'$ corrects for the intensity maintaining the
same phase~$\mean{s_-}=-i|t\alpha+r\beta_-|$. This manifests in the
higher-order correlation functions~(\ref{eq:SunNov19161840CET2017}):
while evaluating them at $\mathcal{F}'=\mathcal{F}_{2,+}'$ does not
lead to small values, for $\mathcal{F}'=\mathcal{F}_{2,-}'$ they
remain close to zero as well (although in general do not recover the
exact zero)~\footnote{For every~$N$ there are two values,
  $\mathcal{F}_{N,\pm}'$, that lead to $\tilde g^{(N)}_a=0$, but they
  do not imply zero values for the other functions in
  general. Remarkably, for a given~$N$ and the parameter
  $\mathcal{F}_{N,-}'$, there is always a $\Gamma_{N,N'}$ for which
  also the coherence function $N'$ is exactly zero, as long as
  $N'>N+1$. Therefore, exact zeros are found for pairs of coherence
  functions, $\{N,N'>N+1\}$, when using this particular pair of
  parameters~$\{\mathcal{F}_{N,-},\Gamma_{N,N'}\}$. For instance,
  $\Gamma_{2,4}=\gamma_\sigma/24$ and
  $\Gamma_{2,5}=(4\pm\sqrt{13})\gamma_\sigma/12$. Since the
  antibunching obtained with our scheme is due to interference between
  coherent and incoherent components and there are only two parameters
  left in Eq.~(\ref{eq:SunNov19163153CET2017}), it is reasonable that
  the condition $\tilde g^{(N)}_a=0$ can be satisfied for two
  different $N$, $N'$ simultaneously, obtaining two conditions for
  $\mathcal{F'}$ and~$\Gamma$.}. Note that, by performing a
\emph{wave-function expansion}, following the procedure in
Refs.~\cite{carmichael91b,bamba11a}, on the joint state of the emitter
and detector, the attenuation fractions in
Eq.~(\ref{eq:WedNov15121440CET2017}) yield a suppression of the
two-photon probability in the
detector~\cite{arxiv_zubizarretacasalengua18a}. This corroborates the
idea that perfect antibunching is recovered thanks to an interference
effect at the two-photon level, that is, involving not only coherently
scattered photons but also the incoherent (second order) ones in
Eq.~(\ref{eq:MonJul21132824CEST2008}).

\begin{figure}[t]
  \centering
  \includegraphics[width=0.7\linewidth]{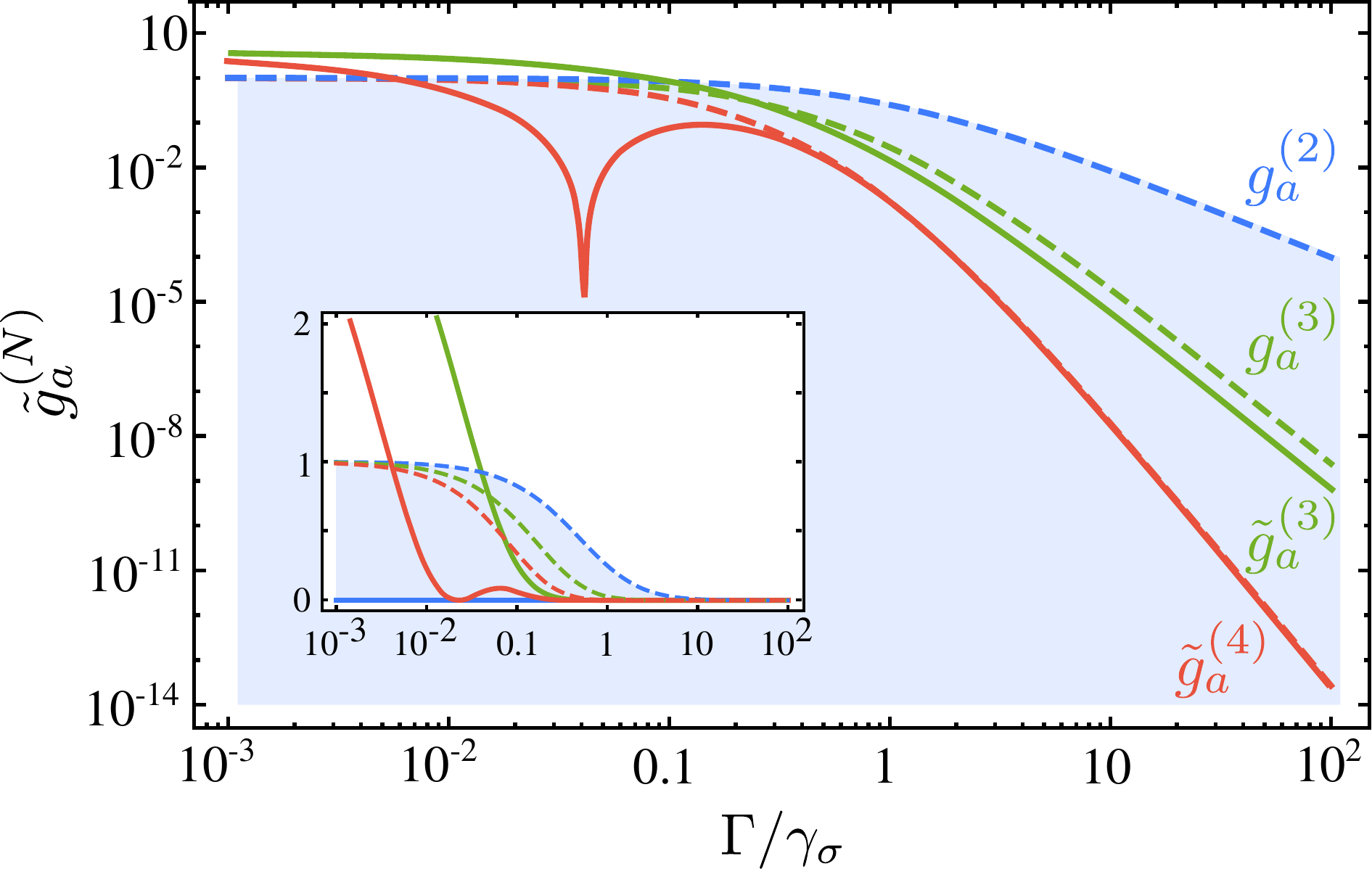}
  \caption{Comparison between, on the one hand, the $N$th-order
    correlation function for the filtered light~$\tilde g_a^{(N)}$
    with the interference signal,
    Eq.~(\ref{eq:SunNov19161840CET2017}), at the condition of perfect
    antibunching, $\mathcal{F}'=\mathcal{F}_{2,-}'$ (solid lines),
    with, on the other hand, the corresponding case $g_a^{(N)}$
    without the interference, i.e., normal resonance fluorescence
    (dashed lines). The main panel is in logarithmic scale and the
    inset in linear scale.  $\tilde g_a^{(2)}$ only appears in the
    linear scale because it is exactly zero for all~$\Gamma$.}
  \label{fig:WedNov15170834CET2017}
\end{figure}

In Fig.~\ref{fig:WedNov15170834CET2017}, we show the correlation
functions for the interference signal,
Eq.~(\ref{eq:SunNov19161840CET2017}), when the condition for perfect
antibunching is met, i.e., $\mathcal{F'}=\mathcal{F}_{2,-}'$, so that
$\tilde g^{(2)}_a=0$ (solid lines), and we compare it to the case
without the interference, i.e., $\mathcal{F}'=0$ (dashed lines), which
is the case from the literature~\cite{nguyen11a,matthiesen12a}. We
plot the cases~$N=2$ (blue), 3~(green) and 4~(red), as a function of
the spectral width of the detector, $\Gamma/\gamma_\sigma$. Note that
the solid blue line does not appear in the main figure which is in
logarithmic scale, because its value is exactly zero to this order in
the Heitler regime. This is the main result as compared to~$g_a^{(2)}$
which, although it can get relatively small, can do so only for broad
linewidths, and loses its antibunching for narrow lines.  In stark
contrast, the perfect antibunching for the interference signal remains
satisfied even when the filter is much narrower than the natural
linewidth of the emitter. On the other hand, the higher-order
correlation functions also yield noteworthy results, which we will
only briefly discuss. In contrast to $\tilde g^{(2)}_a$ which always
remain much smaller than its unfiltered counterpart, there are filter
linewidths where the interference results in larger higher-order
correlations as compared to the standard case. This is clear in the
inset of Fig.~\ref{fig:WedNov15170834CET2017} which is in linear
scale. If one wants to remain within small values of this higher-order
functions, this limits how narrow the filtering can be, though still
allowing for considerable improvement. We have also already noted how
one is limited by the signal. Finally, even though perfect
antibunching remains true for arbitrarily narrow filters in the
theory, as~$\Gamma \ll \gamma_\sigma$, the system would become
unstable under possible small variations of the laser intensity:
$\lim_{\Gamma\rightarrow 0} \mathcal{F}_{2,-}'=2$, which is a
diverging point for $\tilde g^{(2)}_a(\tau)$. A small fluctuation in
the laser intensity would bring the system from perfect antibunching
to a huge superbunching~\cite{arxiv_zubizarretacasalengua18a}. This
could be seen as an advantage, providing a highly tunable quantum
photon source that can be switched between antibunching and bunching
by slightly adjusting the second laser attenuation. However, this
superbunching effect also follows from an interference and is not
linked to $N$-photon emission or other types of structured
emission~\cite{arxiv_zubizarretacasalengua18a}.

A representative filter linewidth for optimal operation can be taken
as~$\Gamma=\gamma_\sigma/5$, which we also use as the reference case
in the following figures, because it is well below the natural emitter
linewidth and brings an improvement essentially everywhere, i.e., we
find that the interference yields the values $\tilde g^{(2)}_a=0$,
$\tilde g^{(3)}_a=0.36$ and $\tilde g^{(4)}_a=0.08$ while without the
interference, one gets $g^{(2)}_a=0.69$, $g^{(3)}_a=0.35$ and
$g^{(4)}_a=0.14$. Note also the existence of a second local minimum
for $\tilde g_a^{(4)}$ in Fig.~\ref{fig:WedNov15170834CET2017}.  We
find that there is such a local minimum for all higher-order
correlators except the one that immediately follows the one that is
exactly cancelled by the interference (i.e., $\tilde g_a^{(3)}$ in
this case). Further discussion of the quantum state generated by this
interference would lead us too far astray, therefore we now turn to
two other quantities of considerable interest for single-photon
emission purposes.

\section{Coherence time and emission rate}
\label{sec:4}

\begin{figure}[t] 
  \centering
  \includegraphics[width=.6\linewidth]{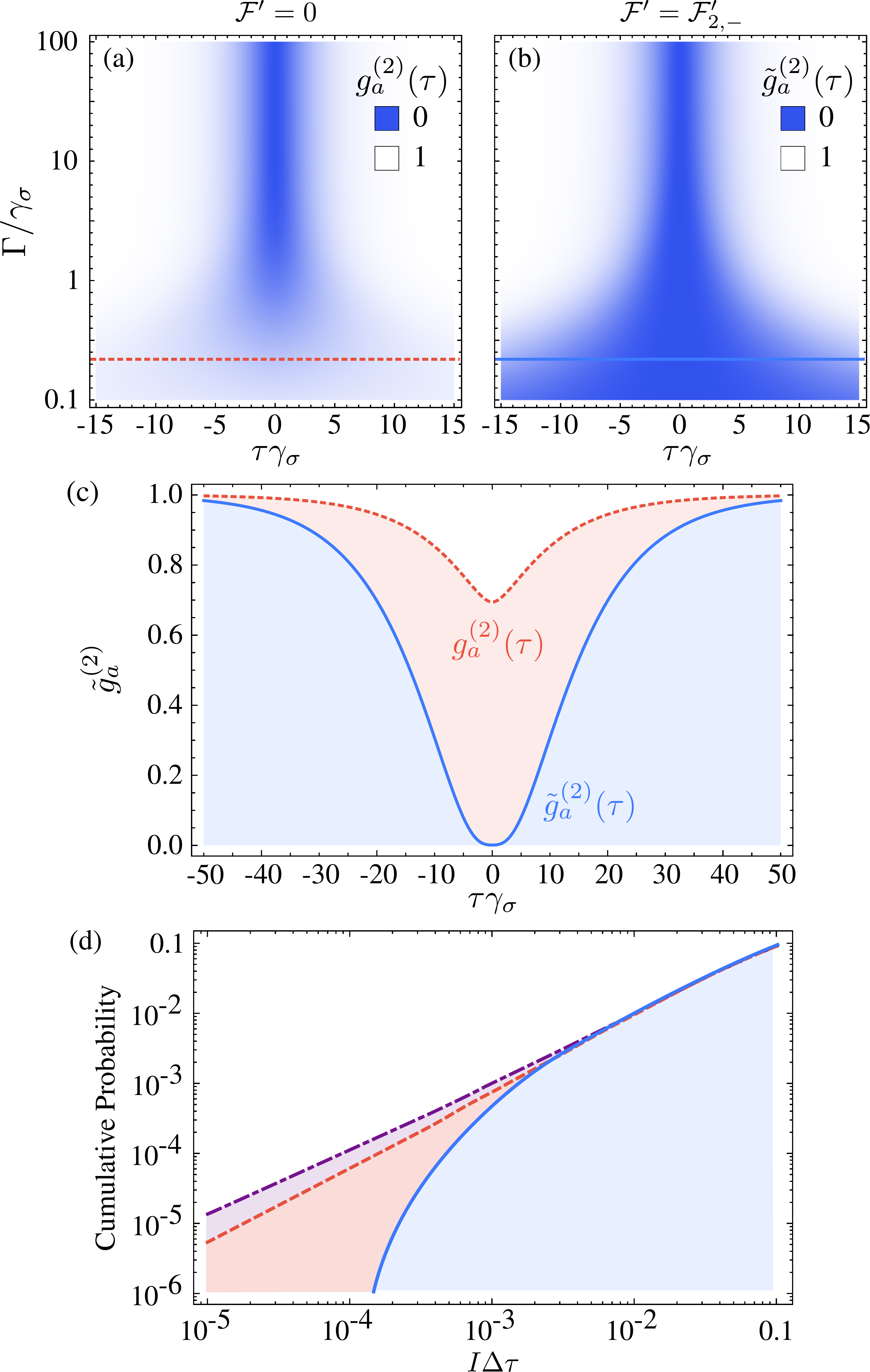}
  \caption{(a, b) Time-dependent second-order correlation function of
    the signal, with ($g_a^{(2)}(\tau)$) and without
    ($\tilde g_a^{(2)}(\tau)$) the interference. Without the
    interference (a), resolving in frequency spoils the
    antibunching. With the interference (b), antibunching remains
    perfect ($\tilde g_a^{(2)}(0)=0$) at all~$\Gamma$, increases its
    coherence time and develops a flat plateau at small time delays.
    (c) Two cuts from the density plots at
    $\Gamma=\gamma_\sigma/5$. The plateau for $\tilde g_a^{(2)}$ is
    not easily distinguished in this scale (it is actually better seen
    in Fig.~\ref{fig:1}(iii)) as extending over $\approx\pm2.5/\gamma_\sigma$
    only.  (d)~Cumulative probability that two consecutive photons are
    emitted with a time separation of \emph{up to}~$\Delta\tau$
    (normalised to the inverse of the emission rate $I$), from a
    coherent or random source (dashed-dotted purple) and from
    resonance fluorescence without~(dashed red) and with~(solid blue)
    the interference from the laser. The case with intererence falls
    much faster and opens a gap of time-separation which photons
    cannot access.}
  \label{fig:ThuNov16115936CET2017}
\end{figure}

So far, we have focused on two aspects of the SPS: its second-order
correlation function at zero-time delay,~$\tilde g^{(2)}_a$ (which by
abuse of language we occasionally refer to as ``antibunching''), and
the spectral width of the emission, as it is observed or,
equivalently, filtered,~$\Gamma$.  There are two other quantities
which are of prime importance to characterise such a source: its
coherence time, which estimates how long the correlations are
retained, and the amount of signal. We now discuss them in turn,
starting with the coherence time, which will show that the best
features of the SPS remain to be presented.

In most single-photon sources, other than the one we present, the
coherence time has to be longer than the temporal resolution of the
detector, or the correlations become randomised. Also, they are
required to evolve smoothly rather than featuring huge oscillations,
that are sometimes observed in the wake of strong
antibunching~\cite{liew10a}.  To characterise our source in this
respect, we consider the time-resolved second-order correlation
function~$\tilde g_a^{(2)}(\tau)$, which can be computed from our
master equation~(\ref{eq:TueNov14183943CET2017}), deriving the
equations from~\ref{app:1} and applying the quantum regression
theorem. Although in the following we present numerical results for
these correlations, so as to easily access arbitrary time delays, the
same procedure as for the zero-delay case can produce some closed-form
but lengthy formulas in the Heitler regime, as is detailed
in~\ref{app:1}~\footnote{$\tilde g_a^{(2)}(\tau)$ could be computed
  analytically in the case of simple detection of resonance
  fluorescence only, with~$\mathcal{F}=0$, using the formulas for the
  frequency and time resolved correlation functions in the
  supplemental material of Ref.~\cite{delvalle12a} and setting the
  detection frequency to zero.}.

In Fig.~\ref{fig:ThuNov16115936CET2017}, we compare the delayed
second-order correlation function as measured by the detectors for
(a)~resonance fluorescence only, i.e.,~setting~$\mathcal{F}'=0$, with
(b)~its interference with the optimally attenuated laser,
i.e.,~$\mathcal{F}'=\mathcal{F}_{2,-}'$, as previously discussed. For
broad enough filters, when~$\Gamma\gg\gamma_\sigma$, the measured
correlations are perfectly antibunched and identical for both
configurations. This happens because such wide filters collect the
full spectrum and the interference occurs naturally. However, as the
width of the filters becomes comparable to the natural linewidth of
the emitter, the behaviour of the correlations in the two
configurations start to differ. Without the interference with the
attenuated laser, antibunching is rapidly lost (cf. dotted red line
in~Fig.~\ref{fig:ThuNov16115936CET2017}(c)) and, in the
limit~$\Gamma\ll\gamma_\sigma$, the emission becomes completely
randomised, with~$\lim_{\Gamma\rightarrow
  0}g^{(2)}_a(\tau)=1$. However, with the interference, perfect
antibunching is preserved regardless of the width of the filter, as
already stated. Here, two new effects are remarkable: i) the coherence
time, or the time between single-photon emission, increases as the
filter width becomes narrower and ii) the correlations display a
plateau of~$\tilde g^{(2)}_a(\tau)=0$ around~$\tau=0$. This plateau is
particularly noteworthy. It is not entirely obvious on the scale of
Fig.~\ref{fig:ThuNov16115936CET2017}(c) since its extent is over
$\tau\gamma_\sigma=\pm2.5$ only, but it results in a dramatic type of
correlations for the photons. Namely, such a plateau, as opposed to
the standard case whose derivative is zero at zero-coincidences only,
corresponds to opening a gap in the time-separation between
consecutive photons, meaning that while the case without interference
makes it only very unlikely to find photon arbitrarily close, the
interference SPS makes it impossible. In this sense, this restores a
notion of ``perfect antibunching'' even though the $\tilde g^{(2)}_a$
to all orders does not cancel exactly. The character of such
correlations is better seen in
Fig.~\ref{fig:ThuNov16115936CET2017}(d), which shows the cumulative
probability that a pair of \emph{consecutive} photons are separated by
a delay of up to~$\Delta\tau$, as a function of this delay which, so
as to compare photon sources with different emission rates, we have
normalised to the mean delay between consecutive photons~(which is
given by the inverse of the emission rate, $1/I$). These results have
been obtained with a Quantum Monte Carlo simulation of the filtered
resonance fluorescence~\cite{lopezcarreno18a}.  We compare three
cases: a coherent (random) source~(dashed-dotted purple), and then the
SPS without~(dashed red) and with~(solid blue) the interference from
the laser. For delays larger than~$I \Delta\tau \approx 10^{-2}$ the
lines for the three sources converge, since at such large delays, the
short-time correlations are lost and dominated by pure randomness,
therefore recovering the uncorrelated case. At about
$I\Delta\tau\approx10^2$, the lines further saturate to unity, as they
should from probability normalisation.

The interesting features lie at short delays. There, it is seen that,
while for the coherent source the cumulative probability increases
linearly as~$I\Delta\tau$ (the exact expression for this simple case
being $1-\exp(I\Delta\tau)$), for the emitter without the
interference, the growth is slower because of its antibunching, which
lowers the probability for photons to be detected close to each
other. The difference is however small and the trend is qualitatively
similar to that of the uncorrelated photon source! Indeed, such a
difference is eclipsed by the type of suppression that is observed by
the emitter with the interference (solid blue line). There, the
departure is much more pronounced and is qualitatively of a different
character, increasing its slope till a point where it would become
vertical, meaning the complete impossibility to ever detect two
photons closer to each other than a finite nonzero time window. This
suppression comes from the plateau in the~$\tilde g^{(2)}_a(\tau)$,
and shows how the enhancement in the correlations that is obtain
through the interference cannot be obtained by using another emitter
operating in a different timescale.

\begin{figure}[t]
  \centering
  \includegraphics[width=.6\linewidth]{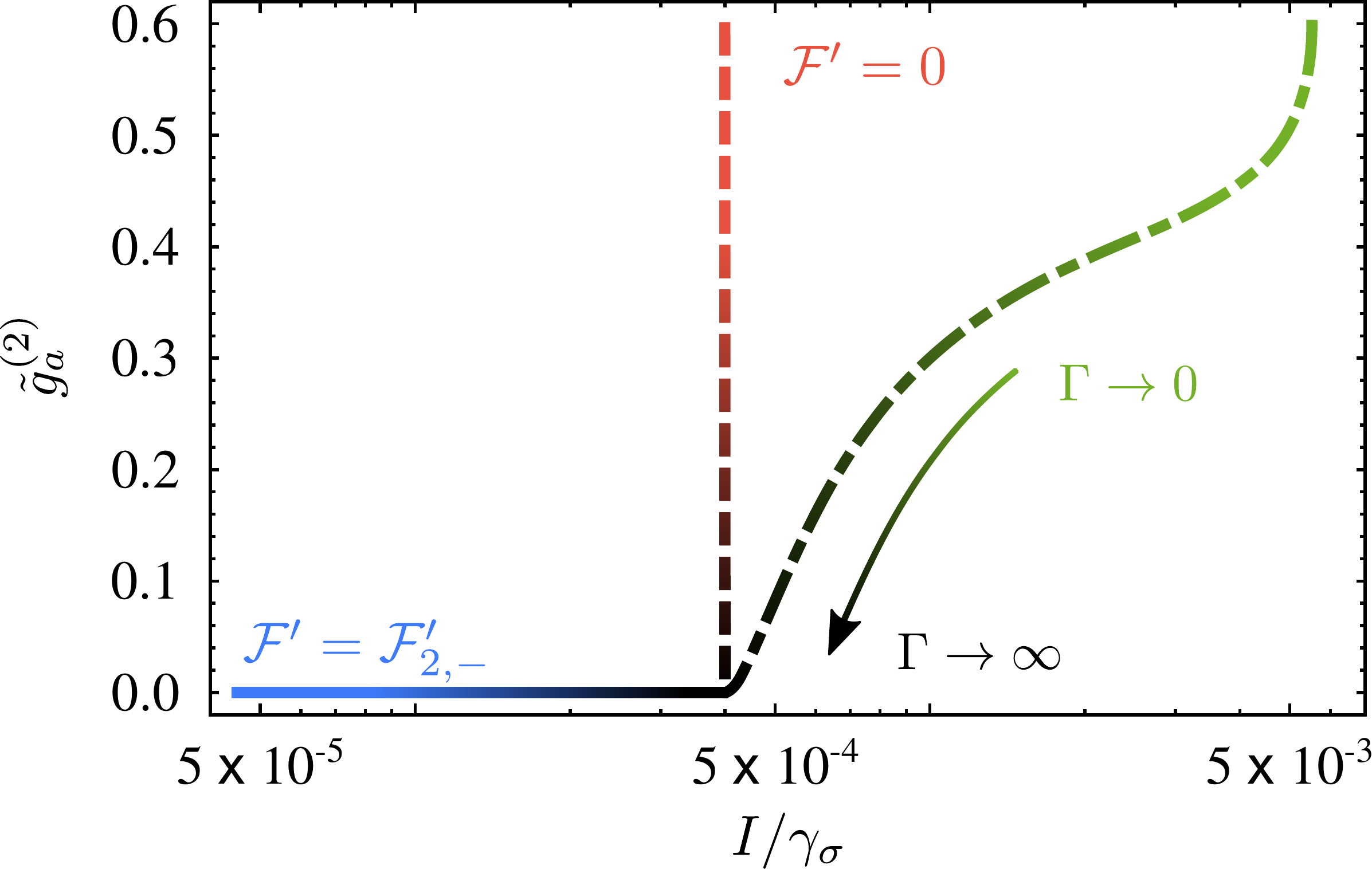}
  \caption{(a) Comparion of the second-order
    correlation~$\tilde g_a^{(2)}$ and emission rate~$I$ for the
    different schemes as a function of the filter linewidth~$\Gamma$
    (which ranges from infinity at the meeting point, in black, to
    zero at the other extremeties of each curve). The dashed red line
    shows the case of filtered resonance fluorescence: the emission
    rate is constant, in agreement with
    Eq.~(\ref{eq:TueJan30143732CET2018}), but as the filter narrows,
    the measured~$g_a^{(2)}$ deviates from 0. The solid blue line
    shows our proposed scheme with the interference (with $t\sim 1$),
    which results in a reduction of the available signal, as shown in
    Eq.~(\ref{eq:TueJan30143620CET2018}), but
    maintaining~$\tilde g_a^{(2)}=0$. The dashed-dotted green line
    shows the case where, for every $\Gamma$, we select a new emitter
    with a different decay rate~$\hat\gamma_\sigma\leq\gamma_\sigma$
    such that, without the interference, the emission spectra and the
    coherence time of the~$g^{(2)}_a(\tau)$ are equal to those
    obtained with our scheme. As the emitters have narrower
    linewidths, the emission rate is larger, but the~$g_a^{(2)}$
    deviates rapidly from zero.}
  \label{fig:WedDec20144623CET2017}
\end{figure}

As a final important characteristic, we have to address the only
feature for which our proposed SPS does not overcome the other types
of sources: the intensity of the signal. One of the acclaimed
qualities of resonance fluorescence is that it is
ultrabright~\cite{nguyen11a,matthiesen12a}, being indeed efficiently
excited by a resonant laser, in the first place. Note that an
incoherently driven SPS would be brighter still as it can saturate the
emitter, providing twice as much signal than under coherent driving
which is limited by stimulated emission. The incoherently pumped SPS
would, however, emit photons of completely undetermined frequency due
to power broadening. In our case, the whole procedure comes at the
price of losing signal, i.e., of reducing the total emission rate
$I=\Gamma \mean{n_a}$.  This is done in two ways: by filtering out the
incoherent fraction or resonance fluorescence to narrow its lineshape
(which technically should also toll other systems claiming subnatural
linewidths) but also, in order to compensate this loss of the
incoherent fraction, by removing part of the coherent fraction through
destructive interferences with the laser.  The total emission will
thus clearly be reduced.  To evaluate the brightness of our homodyne
scheme~$I_\mathrm{int}$, let us first obtain that of filtered
resonance fluorescence~$I_\mathrm{r.f.}$.  The population of the
detector~$\mean{n_a}$ is related to the population of the emitter that
feeds it, $\mean{n_\sigma}$, through the emission spectrum at
resonance~$S_{\Gamma,\sigma}(\omega=0)$ by
\begin{equation}
  \label{eq:FriFeb9112315CET2018}
  \mean{n_a}=\frac{|g|^2}{\Gamma}
  2\pi\mean{n_\sigma}S_{\Gamma,\sigma}(\omega=0)\,,
\end{equation}
(see the equivalences in the supplemental of
Ref.~\cite{delvalle12a}), as long as there is no extra driving
of the detector. Considering the correspondence between the sensor
method~\cite{delvalle12a} and the cascaded formalism~(see
Ref.~\cite{lopezcarreno18a}), we can write
$|g|\rightarrow \sqrt{\Gamma \gamma_\sigma}$. We substitute as well
the spectrum convoluted with the detector, as explained in
Eq.~(\ref{eq:TueDec5115707CET2017}) of the~\ref{app:2}, which
in the Heitler regime is simply
\begin{equation}
  \label{eq:FriFeb9113725CET2018}
  S_{\Gamma,\sigma}(\omega)
=\frac{1}{\pi}\frac{\Gamma/2}{(\Gamma/2)^2+\omega^2}\,.
\end{equation}
Since we are interested in the rates to first order in the
driving~$\Omega_\sigma$, only coherently scattered photons (the first
term in Eq.~(\ref{eq:MonJul21132824CEST2008})) are included in this
derivation. The emission rate from the filtered resonance fluorescence
then converges to the original emission (without detection):
\begin{equation}
  I_\mathrm{r.f.}=\gamma_\sigma\mean{n_\sigma}=
  4\Omega_\sigma^2/\gamma_\sigma\,.
\label{eq:TueJan30143732CET2018}
\end{equation}
In the case of interference with the attenuated laser, the spectrum of
emission remains the same (the coherent part only, to first order
in~$\Omega_\sigma$) but the population is now that of the total
admixed signal:
\begin{equation}
  I_\mathrm{int}=\gamma_\sigma\mean{n_s}\,.
\label{eq:TueJan30143620CET2018}
\end{equation}
This population can be easily computed:
\begin{equation}
  \label{eq:alltogether}
    \mean{n_s}=t^2\mean{n_\sigma}+r^2|\beta|^2+2
                rt\Re[\mean{\sigma}\beta^*]
              =t^2 \left( 1-\frac{\mathcal{F}}{2}\right)^2
      \frac{4\Omega_\sigma^2}{\gamma_\sigma^2}\,.
    \end{equation}
Finally, by comparing Eq.~(\ref{eq:TueJan30143732CET2018}) with
Eqs.~(\ref{eq:TueJan30143620CET2018}--\ref{eq:alltogether}), we find
that the interference with the laser reduces the emission rate by a
factor related to~$\mathcal{F}$, as
\begin{equation}
    \label{eq:TueJan30143832CET2018a}
    \frac{I_\mathrm{int}}{I_\mathrm{r.f.}}=t^2(1-\mathcal{F}/2)^2\,.
\end{equation}
For the condition that yields perfect antibunching
($\mathcal{F}=\mathcal{F}_{2,-}$), this reduces to
\begin{equation}
  \label{eq:TueJan30143832CET2018b}
  \frac{I_\mathrm{int}}{I_\mathrm{r.f.}}=
  t^2 \frac{\Gamma}{\Gamma+\gamma_\sigma}\,. 
\end{equation}
The signal is reduced from the filtering procedure by a factor
${\Gamma}/({\Gamma+\gamma_\sigma})$, times the loss of the
beam-splitter by a factor~$t^2$. This last factor could be overcome by
using an unbalanced beam splitter where $t\sim 1$ and attenuating the
laser accordingly. Still, the brightness is reasonably good, for
instance, for our reference case, $\Gamma=\gamma_\sigma/5$, the
reduction is only of a factor $I_\mathrm{int}/I_\mathrm{r.f.}=1/6$. On
the other hand, we have gained enormously in antibunching and the
linewidth is narrow indeed.

In Fig.~\ref{fig:WedDec20144623CET2017} we can see the comparison
between both cases in a parametric plot where $\Gamma$ is varied, with
the width of the filters being encoded in the color gradient of each
line, starting from black in the limit of~$\Gamma\rightarrow\infty$,
and ending with the respective colors, in the limit
of~$\Gamma\rightarrow 0$. The dashed red line corresponds to the case
of filtered resonance fluorescence but without interference. Its
emission rate remains constant regardless of the filter width
following Eq.~(\ref{eq:TueJan30143732CET2018}), with still most of the
emission being provided by the delta peak anyway. However, $g^{(2)}_a$
is quickly lost, as was previously discussed.  The solid blue line
shows the case with interference, and $\tilde g^{(2)}_a$ there remains
at zero independently of the filter width, but at the cost of lowering
its emission rate, in agreement
with~Eq.~(\ref{eq:TueJan30143620CET2018}).

We compare them with a third case to evidence again that in presence
of the interference, the emitter turns into a qualitatively different
type of SPS. Indeed, one could argue that the problem of finding a
bright monochromatic source with perfect antibunching for a given
detector, could be solved simply by using an emitter with a
linewidth~$\hat\gamma_\sigma$ smaller than that of the detector,
$\hat\gamma_\sigma\ll \Gamma<\gamma_\sigma$, so that, by keeping the
same driving intensity, the emitter can be excited more
efficiently. For the comparison to be fair, the choice
of~$\hat\gamma_\sigma$ would have to be done in such a way that the
properties of the emitted light are equal to the ones obtained with
our scheme, namely: the emission spectrum and the coherence time of
the~$\tilde g^{(2)}_a(\tau)$ have to be comparable in both cases.
This results in the dashed-green line where, for each filter
size~$\Gamma$, we choose an emitter with
linewidth~$\hat\gamma_\sigma<\gamma_\sigma$ such that the emission
spectra and the coherence time are equal to those obtained with our
scheme. With this configuration, the emission rate becomes larger with
broader linewidths (because the emitter with
linewidth~$\hat\gamma_\sigma$ gets excited more easily than the one
with linewidth~$\gamma_\sigma$), but such an enhancement in the
emission rate comes at the prize of an increase in the
zero-delay~$g^{(2)}_a$, as seen in the figure. There, it is clear that
all the sources have a point in common, which is the case when all
frequencies are detected. Imposing some frequency resolution results
in some departures. It appears obvious in the light of the demands
made by a quantum circuit that the only one truely tolerable for
quantum applications are those suffered by our SPS.

We conclude with a quick overview of the feasibility of the proposed
setup.  This type of interference between the original signal and a
controlled laser beam (or \emph{local oscillator}) is a standard
technique in quantum optics, known as \emph{homodyne
  measurement}~\cite{mandel82a,jakeman75a, yuen79a, bondurant84a,
  loudon84a, loudon84b, kitagawa86a, collett87a, xu14a, fischer16a,
  muller16a, dory17a,vogel91a,vogel95a}. In particular, looking at the
second-order correlation function when tuning the laser beam
properties was first suggested by Vogel~\cite{vogel91a,vogel95a}, to
analyse the squeezing properties of the signal. Several recent
works~\cite{muller16a,fischer16a,fischer17a} have also used this
concept with a different objective, namely, to subtract the coherent
fraction from the signal (in their case the emission from a cavity in
strong coupling with a quantum dot). Thanks to this procedure, they
could observe the strong quantum features of the remaining incoherent
fraction such as increased indistinguishability, antibunching or a
pulsed Mollow triplet spectrum. Our analysis, focused
  on the fundamental mechanism, did not include the complications
  present in some of the promising platforms to implement the effect,
  such as quantum dots. There, impact of dephasing for instance should
  be taken into account and we have done this in a follow-up
  work~\cite{arXiv_lopezcarreno18b}. Still other details, like the
  fine-structure splitting, could also be included in increasingly
  more refined studies. On the basis of the physical principle that
  allows the effect to take place, however, we see no a-priori reason
  why it could not fully apply also in more complex structures, at the
  price of possibly heavier expressions for the resonant condition and
  more intricate experimental configurations.  It seems therefore
clear to us that the variations needed to implement our scheme are
definitely within reach of the existing setups and
that one could thus, in this way, finally realise a source with
rapidly vanishing second-order correlation function and subnatural
linewidth. Simultaneously.

\section{Conclusions}
\label{sec:5}

In conclusion, we have shown how to implement a new type of
single-photon source that outperforms what is currently available on
every account except in terms of the available signal. This is based
on a variation of resonance fluorescence in the Heitler regime, which
has been claimed in the literature to provide very good antibunching
as well as spectrally narrow emission. We have shown how such
properties in fact do not coexist in resonance fluorescence in its
bare form, due to neglecting the detection process of the emitted
light, that needs to consider jointly these two properties.

However, it is possible to reach this regime, by compensating for the
loss of antibunching caused by the spectral resolution of the
detector, or, equivalently, by filtering. By decomposing the
second-order correlation function into various types of field
fluctuations, we have shown how one component, the coherent one, can
easily be corrected externally to restore the balance which yields
perfect antibunching (to first order in the driving). We provided an
analytical expression for the condition to fulfil and proposed a setup
to implement this scheme. We find that the light that is produced
indeed provides subnatural linewidth and vanishing antibunching, at
the only cost of a diminished signal, which remains, however, less
than a factor of magnitude drop for reasonable
parameters. Interestingly, the photon correlations in time exhibit a
new qualitative trend, in the form of a plateau, which results in a
time-window where photon-coincidences are suppressed exactly. This
leads to the realization of a perfect SPS, in the sense that a
superconductor is a perfect conductor: our source will never produce a
coincidence in an Hanbury Brown--Twiss setup whose correlation time is
smaller than this plateau. This is true to first order in the driving,
meaning than in an actual setup, the waiting time to observe such a
coincidence will not be infinite, but only as large as required, which
is not possible with a conventional SPS.

\section*{Acknowledgments}

Funding by the Ministry of Science and Education of Russian Federation
(RFMEFI61617X0085), the Universidad Aut\'onoma de Madrid under
contract FPI-UAM 2016, the Spanish MINECO under contract
FIS2015-64951-R (CLAQUE) and the RyC program is gratefully
acknowledged.

\section*{References}

\bibliographystyle{iopart-num}
\bibliography{Sci,books,arXiv} 

\providecommand{\newblock}{}
\begin{thebibliography}{10}
\expandafter\ifx\csname url\endcsname\relax
  \def\url#1{{\tt #1}}\fi
\expandafter\ifx\csname urlprefix\endcsname\relax\def\urlprefix{URL }\fi
\providecommand{\eprint}[2][]{\url{#2}}

\bibitem{heitler_book44a}
Heitler W 1944 {\em The Quantum Theory of Radiation\/} (Oxford University
  Press)

\bibitem{mollow69a}
Mollow B~R 1969 {\em Phys. Rev.\/} {\bf 188} 1969

\bibitem{wu75a}
Wu F~Y, Grove R~E and Ezekiel S 1975 {\em Phys. Rev. Lett.\/} {\bf 35} 1426

\bibitem{agarwal76a}
Agarwal G~S 1976 {\em Phys. Rev. Lett.\/} {\bf 37} 1383

\bibitem{kimble76a}
Kimble H~J and Mandel L 1976 {\em Phys. Rev. A\/} {\bf 13} 2123

\bibitem{cohentannoudji77a}
Cohen-Tannoudji C~N and Reynaud S 1977 {\em J. Phys. B.: At. Mol. Phys.\/} {\bf
  10} 345

\bibitem{grove77a}
Groove R~E, Wu F~Y and Ezekial S 1977 {\em Phys. Rev. A\/} {\bf 16} 227

\bibitem{kimble77a}
Kimble H~J, Dagenais M and Mandel L 1977 {\em Phys. Rev. Lett.\/} {\bf 39} 691

\bibitem{dagenais78a}
Dagenais M and Mandel L 1978 {\em Phys. Rev. A\/} {\bf 18} 2217

\bibitem{knight78a}
Knight P~L, Molander W~A and C~R~Stroud J 1978 {\em Phys. Rev. A\/} {\bf 17}
  1547

\bibitem{apanasevich79a}
Apanasevich P~A and Kilin S~Y 1979 {\em J. Phys. B.: At. Mol. Phys.\/} {\bf 12}
  L83

\bibitem{cohentannoudji79a}
Cohen-Tannoudji C and Reynaud S 1979 {\em Phil. Trans. R. Soc. Lond. A\/} {\bf
  293} 223

\bibitem{mandel79a}
Mandel L 1979 {\em Opt. Lett.\/} {\bf 4} 205

\bibitem{singh83a}
Singh S 1983 {\em Opt. Commun.\/} {\bf 44} 254

\bibitem{carmichael85a}
Carmichael H~J 1985 {\em Phys. Rev. Lett.\/} {\bf 55} 2790

\bibitem{kozlovskii99a}
Kozlovskii A and Oraevskii A 1999 {\em J. Exp. Th. Phys.\/} {\bf 88} 666

\bibitem{astafiev10a}
Astafiev O, Zagoskin A~M, Jr A~A~A, Pashkin Y~A, Yamamoto T, Inomata K,
  Nakamura Y and Tsai J~S 2010 {\em Science\/} {\bf 327} 840

\bibitem{verma11a}
Verma V~B, Stevens M~J, Silverman K~L, Dias N~L, Garg A, Coleman J~J and Mirin
  R~P 2011 {\em Opt. Express\/} {\bf 19} 4182

\bibitem{itano88a}
Berquist W~M~I~J~C and andD J~Wineland 1988 {\em Phys. Rev. A\/} {\bf 38} 559

\bibitem{grangier86b}
Grangier P, Roger G, Aspect A, Heidmann A and Reynaud S 1986 {\em Phys. Rev.
  Lett.\/} {\bf 67} 687

\bibitem{rempe90a}
Rempe G, Schmidt-Kaler F and Walther H 1990 {\em Phys. Rev. Lett.\/} {\bf 64}

\bibitem{diedrich87a}
Diedrich F and Walther H 1987 {\em Phys. Rev. Lett.\/} {\bf 58} 203

\bibitem{bergquist86a}
Bergquist J~C, Hulet R~G, Itano W~M and Wineland D~J 1986 {\em Phys. Rev.
  Lett.\/} {\bf 57} 1699

\bibitem{schubert92a}
Schubert M, Siemers I, Blatt R, Neuhauser W and Toschek P~E 1992 {\em Phys.
  Rev. Lett.\/} {\bf 68} 3016

\bibitem{kask85a}
Kask P, Piksarv P and Mets U 1985 {\em Eur. Biophys. J.\/} {\bf 12} 163

\bibitem{basche92a}
Basch\'{e} T, Moerner W~E, Orrit M and Talon H 1992 {\em Phys. Rev. Lett.\/}
  {\bf 69} 1516

\bibitem{treussart01a}
Treussart F, Clouqueur A, Grossman C and Roch J~F 2001 {\em Opt. Lett.\/} {\bf
  26} 1504

\bibitem{lounis00a}
Lounis B and Moerner W~E 2000 {\em Nature\/} {\bf 407} 491

\bibitem{michler00a}
Michler P, Kiraz A, Becher C, Schoenfeld W~V, Petroff P~M, Zhang L, Hu E and
  \Imamoglu A 2000 {\em Science\/} {\bf 290} 2282

\bibitem{lounis00b}
Lounis B, Bechtel H~A, Gerion D and Moerner P~A~W~E 2000 {\em Chem. Phys.
  Lett.\/} {\bf 329} 399

\bibitem{santori01a}
Santori C, Pelton M, Solomon G, Dale Y and Yamamoto Y 2001 {\em Phys. Rev.
  Lett.\/} {\bf 86} 1502

\bibitem{zwiller02b}
Zwiller V, Blom H, Jonsson P, Panev N, Jeppessen S, Tsegaye T, Goobar E, Pistol
  M~E, Samuelson L and Bj\"ork G 2001 {\em Appl. Phys. Lett.\/} {\bf 78} 2476

\bibitem{sebald02a}
Sebald K, Michler P, Passow T, Hommel D, Bacher G and Forshel A 2002 {\em Appl.
  Phys. Lett.\/} {\bf 81} 2920

\bibitem{santori02a}
Santori C, Fattal D, \Vuckovic J, Solomon G~S and Yamamoto Y 2002 {\em
  Nature\/} {\bf 419} 594

\bibitem{pelton02a}
Pelton M, Santori C, Vu\v{c}kovi\'{c} J, Zhang B, Solomon G, Plant J and
  Yamamoto Y 2002 {\em Phys. Rev. Lett.\/} {\bf 89} 2333602

\bibitem{yuan02a}
Yuan Z, Kardynal B~E, Stevenson R~M, Shields A~J, Lobo C~J, Cooper K, Beattie
  N~S, Ritchie D~A and Pepper M 2002 {\em Science\/} {\bf 295} 102

\bibitem{gerardot05a}
Gerardot B~D, Strauf S, de~Dood M~J~A, Bychkov A~M, Badolato A, Hennessy K, Hu
  E~L, Bouwmeester D and Petroff P~M 2005 {\em Phys. Rev. Lett.\/} {\bf 95}
  137403

\bibitem{bozyigit11a}
Bozyigit D, Lang C, Steffen L, Fink J~M, Eichler C, Baur M, Bianchetti R, Leek
  P~J, Filipp S, da~Silva M~P, Blais A and Wallraff A 2011 {\em Nat. Phys.\/}
  {\bf 7} 154

\bibitem{lang13a}
Lang C, Eichler C, Steffen L, Fink J~M, Woolley M~J, Blais A and Wallraff A
  2013 {\em Nat. Phys.\/} {\bf 9} 345

\bibitem{hoi13a}
Hoi I, Wilson C~M, Johansson G, Lindkvist J, Peropadre B, Palomaki T and
  Delsing P 2013 {\em New J. Phys.\/} {\bf 15} 025011

\bibitem{gu17a}
Gu X, Kochum A~F, Miranowicz C, Liu Y~X and Nori F 2017 {\em Phys. Rep.\/} {\bf
  718-719} 1

\bibitem{kurtsiefer00a}
Kurtsiefer C, Mayer S, Zarda P and Weinfurter H 2000 {\em Phys. Rev. Lett.\/}
  {\bf 85} 290

\bibitem{brouri00a}
Brouri R, Beveratos A, Poizat J~P and Grangier P 2000 {\em Opt. Lett.\/} {\bf
  25} 1294

\bibitem{messin01b}
Messin G, Hermier J~P, Giacobino E, Desbiolles P and Dahan M 2001 {\em Opt.
  Lett.\/} {\bf 23} 1891

\bibitem{kuhlmann15a}
Kuhlmann A~V, Prechtel J~H, Houel J, Ludwig A, Reuter D, Wieck A~D and
  Warburton R~J 2015 {\em Nat. Comm.\/} {\bf 6} 8204

\bibitem{somaschi16a}
Somaschi N, Giesz V, Santis L~D, Loredo J~C, Almeida M~P, Hornecker G,
  Portalupi S~L, Grange T, Ant\'on C, Demory J, G\'omez C, Sagnes I,
  Lanzillotti-Kimura N~D, Lema\^{\i}tre A, Auffeves A, White A~G, Lanco L and
  Senellart P 2016 {\em Nat. Photon.\/} {\bf 10} 340

\bibitem{ding16a}
Ding X, He Y, Duan Z~C, Gregersen N, Chen M~C, Unsleber S, Maier S, Schneider
  C, Kamp M, H\"ofling S, Lu C~Y and Pan J~W 2016 {\em Phys. Rev. Lett.\/} {\bf
  116} 020401

\bibitem{wang16a}
Wang H, Duan Z~C, Li Y~H, Chen S, Li J~P, He Y~M, Chen M~C, He Y, Ding X, Peng
  C~Z, Schneider C, Kamp M, H\"ofling S, Lu C~Y and Pan J~W 2016 {\em Phys.
  Rev. Lett.\/} {\bf 116} 213601

\bibitem{kim16b}
Kim J~H, Cai T, Richardson C~J~K, Leavitt R~P and Waks E 2016 {\em Optica\/}
  {\bf 3} 577

\bibitem{daveau17a}
Daveau R~S, Balram K~C, Pregnolato T, Liu J, Lee E~H, Song J~D, Verma V, Mirin
  R, Nam S~W, Midolo L, Stobbe S, Srinivasan K and Lodahl P 2017 {\em Optica\/}
  {\bf 4} 178

\bibitem{grange17a}
Grange T, Somaschi N, Ant\'{o}n C, Santis L~D, Coppola G, Giesz V, Lema\^{i}tre
  A, Sanges I, Auff\`{e}ves A and Senellart P 2017 {\em Phys. Rev. Lett.\/}
  {\bf 118} 253602

\bibitem{gibbs76a}
Gibbs H~M and Venkatesan T~N~C 1976 {\em Opt. Commun.\/}

\bibitem{hartig76a}
Hartig W, Rasmussen W, Schieder R and Walther H 1976 {\em Z. Phys.\/} {\bf 278}
  205

\bibitem{hoffges97a}
H\"offges J~T, Baldauf H~W, Eichler T, Helmfrid S~R and Walther H 1997 {\em
  Opt. Commun.\/} {\bf 133} 170

\bibitem{nguyen11a}
Nguyen H~S, Sallen G, Voisin C, Roussignol P, Diederichs C and Cassabois G 2011
  {\em Appl. Phys. Lett.\/} {\bf 99} 261904

\bibitem{matthiesen12a}
Matthiesen C, Vamivakas A~N and Atat\"{u}re M 2012 {\em Phys. Rev. Lett.\/}
  {\bf 108} 093602

\bibitem{unsleber16a}
Unsleber S, He Y~M, Gerhardt S, Maier S, Lu C~Y, Pan J~W, Gregersen N, Kamp M,
  Schneider C and H\"ofling S 2016 {\em Opt. Express\/} {\bf 24} 8539

\bibitem{loredo16a}
Loredo J~C, Zakaria N~A, Somaschi N, Anton C, de~Santis L, Giesz V, Grange T,
  Broome M~A, Gazzano O, Coppola G, Sagnes I, Lemaitre A, Auffeves A, Senellart
  P, Almeida M~P and White A~G 2016 {\em Optica\/} {\bf 3} 433

\bibitem{he17a}
He Y~M, Liu J, Maier S, Emmerling M, Gerhardt S, Davan\c{c}o M, Schneider C and
  H\"{o}fling S 2017 {\em Optica\/} {\bf 4} 802

\bibitem{gonzaleztudela13a}
Gonz\'alez-Tudela A, Laussy F~P, Tejedor C, Hartmann M~J and del Valle E 2013
  {\em New J. Phys.\/} {\bf 15} 033036

\bibitem{eberly77a}
Eberly J and W\'odkiewicz K 1977 {\em J. Opt. Soc. Am.\/} {\bf 67} 1252

\bibitem{knoll86a}
Kn\"oll L and Weber G 1986 {\em J. Phys. B.: At. Mol. Phys.\/} {\bf 19} 2817

\bibitem{knoll90a}
Kn\"oll L, Vogel W and Welsch D~G 1990 {\em Phys. Rev. A\/} {\bf 42} 503

\bibitem{delvalle12a}
del Valle E, Gonz\'alez-Tudela A, Laussy F~P, Tejedor C and Hartmann M~J 2012
  {\em Phys. Rev. Lett.\/} {\bf 109} 183601

\bibitem{gardiner_book00a}
Gardiner G~W and Zoller P 2000 {\em Quantum Noise\/} 2nd ed (Springer-Verlag,
  Berlin)

\bibitem{lopezcarreno18a}
{L\'{o}pez Carre\~{n}o} J~C, {del Valle} E and Laussy F~P 2018 {\em Sci.
  Rep.\/} {\bf 8} 6975

\bibitem{glauber63a}
Glauber R~J 1963 {\em Phys. Rev. Lett.\/} {\bf 10} 84

\bibitem{loudon_book00a}
Loudon R 2000 {\em The quantum theory of light\/} 3rd ed (Oxford Science
  Publications)

\bibitem{dalibard83a}
Dalibard J and Reynaud S 1983 {\em J. Phys. France\/} {\bf 44} 1337

\bibitem{lopezcarreno16a}
{L\'opez Carre{\~n}o} J~C and Laussy F~P 2016 {\em Phys. Rev. A\/} {\bf 94}
  063825

\bibitem{mandel82a}
Mandel L 1982 {\em Phys. Rev. Lett.\/} {\bf 49} 136

\bibitem{vogel91a}
Vogel W 1991 {\em Phys. Rev. Lett.\/} {\bf 67} 2450

\bibitem{vogel95a}
Vogel W 1995 {\em Phys. Rev. A\/} {\bf 51} 4160

\bibitem{slusher85a}
Slusher R~E, Hollberg L~W, Yurke B, Mertz J~C and Valley J~F 1985 {\em Phys.
  Rev. Lett.\/} {\bf 55} 2409

\bibitem{mccormick07a}
McCormick C~F, Boyer V, Arimondo E and Lett P~D 2007 {\em Opt. Lett.\/} {\bf
  32} 178

\bibitem{mccormick08a}
McCormick C~F, Marino A~M, Boyer V and Lett P~D 2008 {\em Phys. Rev. A\/} {\bf
  78} 043816

\bibitem{raizen87a}
Raizen M~G, Orozco L~A, Xiao M, Boyd T~L and Kimble H~J 1987 {\em Phys. Rev.
  Lett.\/} {\bf 69} 198

\bibitem{lu98a}
Lu Z~H, Bali S and Thomas J~E 1998 {\em Phys. Rev. Lett.\/} {\bf 81} 3635

\bibitem{ourjoumtsev11a}
Ourjoumtsev A, Kubanek A, Koch M, Sames C, Pinkse P~W~H, Rempe G and Murr K
  2011 {\em Nature\/} {\bf 474} 623

\bibitem{schulte15a}
Schulte C~H~H, Hansom J, Jones A~E, Matthiesen C, Gall C~L and Atat\"{u}re M
  2015 {\em Nature\/} {\bf 525} 222

\bibitem{arxiv_zubizarretacasalengua18a}
{Zubizarreta Casalengua} E, {L{\'o}pez Carre{\~n}o} J~C, Laussy F~P and del
  Valle E 2018 {\em In preparation\/}

\bibitem{zubizarretacasalengua17a}
{Zubizarreta Casalengua} E, {L\'opez Carre\~no} J~C, del Valle E and Laussy F~P
  2017 {\em J. Math. Phys.\/} {\bf 58} 062109

\bibitem{carmichael91b}
Carmichael H~J, Brecha R~J and Rice P~R 1991 {\em Opt. Commun.\/} {\bf 82} 73

\bibitem{bamba11a}
Bamba M, \Imamoglu A, Carusotto I and Ciuti C 2011 {\em Phys. Rev. A\/} {\bf
  83} 021802(R)

\bibitem{liew10a}
Liew T~C~H and Savona V 2010 {\em Phys. Rev. Lett.\/} {\bf 104} 183601

\bibitem{jakeman75a}
Jakeman E, Oliver C~J and Pike E~R 1975 {\em Adv. Phys.\/} {\bf 24} 349

\bibitem{yuen79a}
Yuen H~P and Shapiro J~H 1979 {\em Opt. Lett.\/} {\bf 4} 334

\bibitem{bondurant84a}
Bondurant R~S and Shapiro J~H 1984 {\em Phys. Rev. D\/} {\bf 30} 2548

\bibitem{loudon84a}
Loudon R 1984 {\em Opt. Commun.\/} {\bf 49} 24

\bibitem{loudon84b}
Loudon R 1984 {\em Opt. Commun.\/} {\bf 49} 67

\bibitem{kitagawa86a}
Kitagawa M and Yamamoto Y 1986 {\em Phys. Rev. A\/} {\bf 34} 3974

\bibitem{collett87a}
Collett M~J, Loudon R and Gardiner C~W 1987 {\em J. Mod. Opt.\/} {\bf 34} 881

\bibitem{xu14a}
Xu X~W and Li Y 2014 {\em Phys. Rev. A\/} {\bf 90} 033832

\bibitem{fischer16a}
Fischer K~A, M\"{u}ller K, Rundquist A, Sarmiento T, Piggott A~Y, Kelaita Y~A,
  Dory C, Lagoudakis K~G, M\"uller K and \Vuckovic J 2016 {\em Nat. Photon.\/}
  {\bf 10} 163

\bibitem{muller16a}
M\"{u}ller K, Fischer K~A, Dory C, Sarmiento T, Lagoudakis K~G, Rundquist A,
  Kelaita Y~A and Vu\v{c}kovi\'{c} J 2016 {\em Optica\/} {\bf 3} 931

\bibitem{dory17a}
Dory C, Fischer K~A, M\"uller K, Lagoudakis K~G, Sarmiento T, Rundquist A,
  Zhang J~L, Kelaita Y, Sapra N~V and \Vuckovic J 2017 {\em Phys. Rev. A\/}
  {\bf 95} 023804

\bibitem{fischer17a}
Fischer K~A, Kelaita Y~A, Sapra N~V, Dory C, Lagoudakis K~G, M\"uller K and
  \Vuckovic J 2017 {\em Phys. Rev. Appl.\/} {\bf 7} 044002

\bibitem{arXiv_lopezcarreno18b}
{L{\'o}pez Carre{\~n}o} J~C, {Zubizarreta Casalengua} E, Laussy F~P and del
  Valle E 2018 {\em arXiv:1806.08774\/}

\bibitem{delvalle09a}
del Valle E, Laussy F~P and Tejedor C 2009 {\em Phys. Rev. B\/} {\bf 79} 235326

\end{thebibliography}


 \pagebreak
\appendix
\setcounter{equation}{0}
\setcounter{figure}{0}

\section*{Appendices}

\section{Steady state of the combined resonance fluorescence and
  detector at vanishing laser driving}
\label{app:1}

We first solve the dynamics for the mean value of any system operator,
which in its most general normally ordered form
reads~\cite{delvalle09a}~$C_{\{m,n,\mu,\nu\}}=\mean{\sigma^{\dagger
    m}\sigma^n a^{\dagger \mu} a^\nu}$ (with~$m$, $n \in\{0,1\}$ and
$\mu$, $\nu\in \mathbb{N}$). It follows the equation:
\begin{equation}
  \label{eq:TueMay5174356GMT2009}
  \partial_t
  C_{\{m,n,\mu,\nu\}}=\sum_{{m',n',\mu',\nu'}}\mathcal{M}_{{m,n,\mu,\nu\atop
      m',n',\mu',\nu'}}C_{\{m',n',\mu',\nu'\}}\,,
\end{equation}
with the regression matrix
elements~$\mathcal{M}_{{m,n,\mu,\nu\atop m',n',\mu',\nu'}}$ given by,
in our case:
%
\begin{eqnarray}
  \label{eq:TueMar6173803CET2018a}
\fl    &\mathcal{M}_{{m,n,\mu,\nu\atop
      m,n,\mu,\nu}}=-\frac{\gamma_\sigma}2(m+n)-\frac{\Gamma}2(\mu+\nu)\,,
      \quad&\mathcal{M}_{{m,n,\mu,\nu\atop
              m,1-n,\mu,\nu}}=-i\Omega_\sigma[n+2m(1-n)]\,\\
\fl    &\mathcal{M}_{{m,n,\mu,\nu\atop 1-m,n,\mu,\nu}}=
      i\Omega_\sigma[m+2n(1-m)]\,,\quad
           &\mathcal{M}_{{m,n,\mu,\nu\atop m,n,\mu,\nu-1}}=\Omega_a\nu\,,\\
 \fl   &\mathcal{M}_{{m,n,\mu,\nu\atop m,n,\mu-1,\nu}}=\Omega_a\mu\,,
      \quad &\mathcal{M}_{{m,n,\mu,\nu\atop
              1-m,n,\mu-1,\nu}}=-g(1-m)\mu\,,\\
  \label{eq:TueMar6173803CET2018d}
\fl    &\mathcal{M}_{{m,n,\mu,\nu\atop m,1-n,\mu,\nu-1}}=-g(1-n)\nu
  \end{eqnarray}
%
and zero everywhere else. These equations can be solved numerically,
choosing a high enough truncation in the number of photons, in order
to obtain a converged steady state
($\partial_t C_{\{m,n,\mu,\nu\}}=0$) for any given pump power.
However, it is possible to derive analytical solutions in the case
where we use a ``sensor'' ($g\rightarrow 0$) in the vanishing driving
limit ($\Omega_\sigma \rightarrow 0$ after setting
$\Omega_a=g\Omega_\sigma \mathcal{F}/\gamma_\sigma$). In this case, it
is enough to solve recursively sets of truncated equations. That is,
we start with the lowest order correlators, with only one operator,
which we write in a vectorial form for convenience:
$\mathbf{v}_1=(\mean{a}, \mean{a^\dagger}, \mean{\sigma},
\mean{\sigma^\dagger })^\mathrm{T}$. Its equation reads
$\partial_t \mathbf{v}_1= M_1
\mathbf{v}_1+A_1+o(\Omega,g)$ 
where $o(\Omega,t)$ means higher-order terms of these variables,
where~$\Omega$ stands for both $\Omega_a$ and~$\Omega_\sigma$.  This
provides the steady state
value~$\mathbf{v}_1= -M_1^{-1} A_1+o(\Omega,g)$. We proceed in the
same way with the two-operator correlators
$\mathbf{v}_2=( \mean{a^2}, \mean{a^{\dagger 2}}, \mean{a^\dagger a},
\mean{\sigma^\dagger\sigma}, \mean{\sigma^\dagger
  a},\,\cdots)^\mathrm{T}$, only, in this case, we also need to
include the steady state value for the one-operator correlators as
part of the independent term in the equation:
$\partial_t \mathbf{v}_2= M_2 \mathbf{v}_2+A_2+X_{21}
\mathbf{v}_1+o(\Omega,t)$. The steady state
reads~$\mathbf{v}_2= -M_2^{-1} (A_2+X_{21} \mathbf{v}_1)+o(\Omega,g)$
with a straightforward generalisation
$\mathbf{v}_N= -M_N^{-1} (A_N+\sum_{j=1}^{N-1}X_{N j}
\mathbf{v}_j)+o(\Omega,g)$.

We are interested in this text in photon correlators of the
form~$\mean{a^{\dagger N} a^N }$. These
follow~$\mean{a^{\dagger N} a^N }\sim (\Omega_\sigma g)^{2N}$, to
lowest order in both $\Omega_\sigma$ and~$g$. The normalised
correlation functions~$g_\Gamma^{(N)}$ are thus independent of both
$\Omega_\sigma$ and~$g$ to lowest order, and their computation
requires to solve the $2N$ sets of recurrent equations and taking the
limits~$\lim_{g\rightarrow 0}\lim_{\Omega_\sigma\rightarrow
  0}\mean{a^{\dagger N} a^N }/\mean{a^\dagger a}^N $. This can be done
analytically and this provides Eqs.~(\ref{eq:WedNov15091914CET2017})
and (\ref{eq:SunNov19161840CET2017}) from the main text.

\section{Two-time correlators and spectrum of emission for resonance fluorescence (at any laser driving, without detector)}
\label{app:2}

First, using again the regression
matrix~(\ref{eq:TueMar6173803CET2018a}-\ref{eq:TueMar6173803CET2018d}),
we write the equations~(\ref{eq:TueMay5174356GMT2009}) in a vectorial
form for the two-level system only, by setting $g=0$.  In this case,
one-time correlators follow
$\partial_\tau\mathbf{w}[1,1](\tau)=M_\sigma\mathbf{w}[1,1](\tau)+A_\sigma$
with
\begin{equation}\fl
  \label{eq:TueDec23120004CET2008}
  \mathbf{w}[1,1](\tau)= \left(
  \matrix{
    \mean{\sigma}(\tau)\cr
    \mean{\sigma^\dagger }(\tau)\cr
    \mean{ \sigma^\dagger \sigma}(\tau)\cr
}\right)
\,,\quad
A_\sigma=i\Omega_\sigma \left(
  \matrix{
    -1\cr
    1\cr
    0\cr
  }\right)
  \,,\quad
  M_\sigma=\left (
    \matrix{
      -\frac{\gamma_\sigma}{2} & 0 &2i\Omega_\sigma\cr
      0 &-\frac{\gamma_\sigma}{2} &-2i\Omega_\sigma\cr
      i\Omega_\sigma&-i\Omega_\sigma&-\gamma_\sigma
}\right) \,.
\end{equation}
The steady state solution reads
\begin{equation}
  \label{eq:TueNov21003718CET2017}
  \mathbf{w}[1,1]= \left(
  \matrix{
    \mean{\sigma}\cr
    \mean{\sigma^\dagger }\cr
    \mean{ \sigma^\dagger \sigma}\cr } \right)
=-M_\sigma^{-1}A_\sigma=
  \frac{2\Omega_\sigma}{\gamma_\sigma^2+8\Omega_\sigma^2} 
\left(  \matrix{
    -i \gamma_\sigma\cr
    i \gamma_\sigma\cr
    \Omega_\sigma\cr
} \right) \,.
\end{equation}
By applying the quantum regression theorem which states that two-time
correlators follow the same equations for the time-delay as the
single-time ones for time, we have that
$\partial_\tau\mathbf{w}[L,R](\tau)=M_\sigma
\mathbf{w}[L,R](\tau)+A_\sigma\mean{LR}$ for any two operators $L$,
$R$, with
\begin{equation}
  \label{eq:TueNov21003107CET2017}
  \mathbf{w}[L,R](\tau)= \left(
  \matrix{
    \mean{L\, \sigma(\tau)R}\cr
    \mean{L\,\sigma^\dagger (\tau)R}\cr
    \mean{L \,(\sigma^\dagger \sigma) (\tau)R}\cr
}\right)
\end{equation}
and $\mathbf{w}[L,R](0)$ obtained from the single-time mean values
in~$\mathbf{w}[1,1]$. The solution is given by:
\begin{eqnarray}
  \label{eq:TueNov30133208CET2010}
\fl  \mathbf{w}[L,R](\tau)&=&e^{M_\sigma\tau}\left\{
  \mathbf{w}[L,R](0)+M_\sigma^{-1}A_\sigma\mean{LR}\right\}-M_\sigma^{-1}A_\sigma\mean{LR}\nonumber 
  \\
\fl  &=&e^{M_\sigma\tau}\left\{\mathbf{w}[L,R](0)-\mathbf{w}[1,1]\mean{LR}\right\}+\mathbf{w}[1,1]\mean{LR}\,.
\end{eqnarray}

We compute the correlators that we need below and in the main text, by
solving only two of these two-time correlator vectors, for
$\mathbf{w}[\sigma^\dagger ,\sigma](\tau)$ and
$\mathbf{w}[1,\sigma](\tau)$, since we have
\begin{eqnarray}
  \label{eq:TueNov28114504CET2017}
\fl    &\mean{\sigma^\dagger (\sigma^\dagger
      \sigma)(\tau)\sigma}=\mathbf{w}[\sigma^\dagger
      ,\sigma]_3(\tau)\,,\quad
    &\mean{\sigma^\dagger \sigma^\dagger
      (\tau)\sigma}=\mathbf{w}[\sigma^\dagger ,\sigma]_2(\tau)\,,\nonumber\\ 
\fl    &\mean{(\sigma^\dagger
      \sigma)(\tau)\sigma}=\mathbf{w}[1,\sigma]_3(\tau)\,,\quad
    &\mean{\sigma^\dagger (\tau)\sigma}=\mean{\sigma^\dagger
      \sigma(\tau)}^*=\mathbf{w}[1,\sigma]_2(\tau)\,,\\
\fl    &\mean{\sigma^\dagger \sigma^\dagger (\tau)}=\mean{\sigma(\tau)\sigma}^*=\left\{\mathbf{w}[1,\sigma]_1(\tau)\right\}^*\,.\nonumber
\end{eqnarray}
The initial conditions read
\begin{equation}
  \label{eq:TueNov21005827CET2017}
  \mathbf{w}[\sigma^\dagger ,\sigma](0)= \left(
  \matrix{
    0\cr
    0\cr
    0\cr
  }\right)
  \quad\mathrm{and}\quad
  \mathbf{w}[1,\sigma](0)=\left(
  \matrix{
    0\cr
    \mean{\sigma^\dagger \sigma}\cr
    0\cr
  } \right) \,.  
\end{equation}

We can thus provide the expression for the second-order correlation
function of resonance fluorescence with perfect time resolution (or
without coupling to a detector):
\begin{equation}
  \label{eq:ThuNov23114958CET2017}
 g^{(2)}_\sigma (\tau) = 1 - \Big[\frac{3\gamma_\sigma}{4R_\sigma}
   \sinh(R_\sigma\tau) + \cosh(R_\sigma\tau)\Big] e^{-3\gamma_\sigma
   \tau/4}\,,
\end{equation}
in terms of
$R_\sigma=\sqrt{({\gamma_\sigma}/{4})^2-(2\Omega_\sigma)^2}$. Note
that $\lim_{\Omega_\sigma\rightarrow 0} R_\sigma =\gamma_\sigma/4$ and
that oscillations only appear in $g^{(2)}_\sigma (\tau)$ when
$\Omega_\sigma>\gamma_\sigma/8$, as the two-level system enters into
strong coupling with the laser.

On the other hand, the normalised steady state spectrum of emission
with perfect frequency resolution~\cite{mollow69a,loudon_book00a}, is
defined as
\begin{equation} 
\label{eq:TueNov28145129CET2017}
S_\sigma(\omega)=\frac{1}{\pi
  \mean{n_\sigma}}\Re\int_{0}^{\infty}\mean{\sigma^\dagger
  \sigma(\tau)}e^{i\omega\tau}d\tau\,.
\end{equation} 
Substituting the expression found for the
correlator~$\mean{\sigma^\dagger \sigma(\tau)}$ in
Eq.~(\ref{eq:TueNov28114504CET2017}) and expanding up to second order
in the driving~$\Omega_\sigma$, we obtain the
formula~(\ref{eq:MonJul21132824CEST2008}) in the main text.

The expression after convolution with a detector with spectral
resolution $\Gamma$ is:
\begin{equation} 
\label{eq:TueDec5115707CET2017}
S_{\Gamma,\sigma}(\omega)=\frac{1}{\pi
  \mean{n_\sigma}}\Re\int_{0}^{\infty}\mean{\sigma^\dagger
  \sigma(\tau)}e^{(i\omega-\Gamma/2)\tau}d\tau\,,
\end{equation} 
which is used to obtain Eq.~(\ref{eq:FriFeb9113725CET2018}).

\end{document}